\documentclass{pasa}%
\usepackage{graphicx}
\usepackage{caption}
\usepackage{subcaption}

\title[Radio halos revealed by ASKAP]{ASKAP reveals giant radio halos in two merging SPT galaxy clusters \\[0.2em]\Large{-- Making the case for a direction-dependent pipeline --}}
\author[A. G. Wilber et al]{A. G. Wilber$^{1, 2}$\thanks{Email: amanda.wilber@curtin.edu.au}, M. Johnston-Hollitt$^1$, S. W. Duchesne$^1$, C. Tasse$^3$, H. Akamatsu$^4$, H. Intema$^1$, and T. Hodgson$^1$\\

\affil{$^1$Curtin Institute of Radio Astronomy, 1 Turner Avenue, Technology Park, Bentley WA 6102, Australia}%
\affil{$^2$Hamburger Sternwarte, Universit{\"a}t Hamburg, Gojenbergsweg  112, 21029 Hamburg, Germany}
\affil{$^3$GEPI, Observatoire de Paris, Universite PSL, CNRS, 5 place Jules Janssen, 92190 Meudon, France}
\affil{$^4$SRON Netherlands Institute for Space Research, Utrecht, The Netherlands}
}%

\jid{PASA}
\doi{10.1017/pas.\the\year.xxx}
\jyear{\the\year}

\usepackage{aas_macros}
\usepackage{hyperref} 
\hypersetup{colorlinks,citecolor=blue,linkcolor=blue,urlcolor=blue}

\begin{document}

\begin{frontmatter}
\maketitle

\begin{abstract}
Early science observations from the Australian Square Kilometre Array Pathfinder (ASKAP) have revealed clear signals of diffuse radio emission associated with two clusters detected by the South Pole Telescope via their Sunyaev Zel'dovich signal. SPT CLJ0553-3342 (MACSJ0553.4-3342) and SPT CLJ0638-5358 (Abell S0592) are both high-mass lensing clusters that have undergone major mergers. To improve the data products of these ASKAP early science observations and create science-fidelity images of the galaxy clusters, we performed direction-dependent (DD) calibration and imaging using state-of-the-art software {\sc killMS} and {\sc DDFacet}. We find that artefacts in the ASKAP images are greatly reduced after directional calibration. Here we present our DD calibrated ASKAP radio images of both clusters showing unambiguous giant radio halos with largest linear scales of $\sim1$~Mpc. The halo in MACSJ0553.4-3342 was previously detected with GMRT observations at 323 MHz, but appears more extended in our ASKAP image. Although there is a shock detected in the thermal X-ray emission of this cluster, we find that the particle number density in the shocked region is too low to allow for the generation of a radio shock. The radio halo in Abell S0592 is a new discovery, and the Southwest border of the halo coincides with a shock detected in X-rays. We discuss the origins of these halos considering both the hadronic and turbulent re-acceleration models as well as sources of \textit{seed} electrons. This work gives a positive indication of the potential of ASKAP's Evolutionary Map of the Universe (EMU) survey in detecting intracluster medium radio sources, and showcases the improvement in data products after utilising third-generation calibration techniques.
\end{abstract}

\end{frontmatter}

\section{INTRODUCTION }
\label{sec:intro}

Galaxy clusters are Megaparsec-sized systems that contain hundreds of individual galaxies which reside within a hot pool of ionized, magnetized gas. While some resident galaxies are typically observed to emit radio emission from their active galactic nuclei (AGN), very extended and diffuse radio sources originating from the intracluster medium (ICM) have been identified in a growing number of galaxy clusters over the last several decades \citep[see][for a recent review]{2019SSRv..215...16V}. Theoretical and observational evidence support the suggestion that diffuse radio sources associated with the ICM are generated from the collisions, or mergers, of multiple galaxy clusters \citep[e.g.][]{2001ApJ...553L..15B, 2010ApJ...721L..82C}. As dictated by the evolution of large-scale structure in the Universe, a cluster-cluster merger occurs on a timescale of about 1 Gyr \citep{2002ASSL..272....1S}, and during this time shock waves and turbulence are magnetically driven throughout the system. Turbulence and shocks can excite -- and re-excite -- relativistic electrons within the cluster magnetic field, leading to observable synchrotron emission at radio frequencies \citep[e.g.][]{2008Natur.455..944B, 2012SSRv..166..187B, 2014IJMPD..2330007B, 2015ASSL..407..599B}.

Merger-induced synchrotron sources can come in various forms \citep[see][for a taxonomy]{2004rcfg.proc..335K}. Some merging clusters host diffuse emission, known as \textit{giant radio halos}, which fill the inner volume of the ICM. Other clusters show large, elongated structures on the outer edges of the ICM, called gischt-type \textit{radio relics} or \textit{radio shocks}. There are also cases where extended emission from resident radio galaxies is observed to be shock compressed or re-energised by a merger event of the host cluster. This type of emission does not always fit into a specific classification but examples include the radio phoenixes in A13, A85, A133, and A4038 \citep{2001AJ....122.1172S}, the gently re-energised tail (GReET) in Abell 1033 \citep{2017SciA....3E1634D}, the re-brightened tail in Abell 1132 \citep{2018MNRAS.473.3536W}, and the revived fossil plasma source found in Abell 1314 (\citealp{2019A&A...622A..25W}, van Weeren et al., in prep). 

Radio halos are widely considered to be the result of turbulent re-acceleration of mildly-relativistic electrons in the ICM \citep[e.g.][]{2001MNRAS.320..365B, 2013MNRAS.429.3564D, 2017MNRAS.465.4800P}. However, theory indicates that there should also be a population of secondary radio-emitting CRe produced from collisions between thermal protons and cosmic-ray protons in the ICM \citep[hadronic model;][]{1980ApJ...239L..93D}. It is still unknown as to how much of a role these secondary CRe play in generating radio halos since the expected gamma-ray contribution from these proton-proton collisions has yet to be detected in a single galaxy cluster \citep[e.g.][]{2010ApJ...717L..71A, 2014ApJ...787...18A, 2014A&A...567A..93P}.

Gischt-type relics, or radio shocks, are often found to trace bow shocks occurring on the largest scales, and are therefore thought to be connected to diffusive shock acceleration (DSA) that operates at a shock front \citep{1978ApJ...221L..29B, 1998A&A...332..395E}. X-ray observations of merging clusters typically show an overall disturbed morphology in their thermal gas, however, shock-heated gas can be identified by a sharp discontinuity in surface brightness and a corresponding jump from higher temperature to lower temperature for post-shock and pre-shock regions, respectively. To add to the challenge of their interpretation, not all merging clusters show evidence of shocks and not all detected shocks in merging clusters have radio counterparts \citep[e.g.][]{2018MNRAS.476.3415W, 2018MNRAS.476.5591B}.

The details of the necessary acceleration mechanisms and the efficiencies of low Mach number shocks are still under scientific scrutiny \citep[see][for a recent study]{2020A&A...634A..64B}. Fermi-II acceleration prompted by merger turbulence and Fermi-I acceleration prompted by merger shocks are both usually too weak to accelerate electrons from thermal to ultra-relativistic energies \citep{2005ApJ...627..733M, 2008ApJ...682..175P, 2012ApJ...756...97K, 2016ApJ...818..204V,  2016MNRAS.460L..84B, 2016MNRAS.461.1302E}. Hence, some form of pre-acceleration or a population of relativistic \textit{seed} electrons is required to explain cluster-scale halos and relics. While primordial accretion shocks are suspected to account for a portion of the CRe population in clusters, AGN are another viable source for seed electrons. Several examples have been found where extended radio galaxies appear to be feeding into diffuse cluster sources: [e. g.] the remnant radio galaxy supplying seeds for the relic in the Bullet cluster 1E 0657-55.8 \citep{2015MNRAS.449.1486S}; the connection between a head-tail radio galaxy and relic in Abell 3411-3412 \citep{2017NatAs...1E...5V, 2017NatAs...1E..14J}; and the connection between the giant radio galaxy and the ultra-steep halo in Abell 1132 \citep{2018MNRAS.473.3536W}.

The origins of cluster magnetic fields are predicted to come from a constant primordial field that has been amplified over time \citep[see][for reviews]{2001PhR...348..163G, 2018SSRv..214..122D}. Recent investigations of the evolution of magnetic fields in merging galaxy clusters -- through high-resolution cosmological magnetohydrodynamical simulations -- have revealed that although major mergers can shift the peak magnetic spectra to small scales (via small-scale dynamo) in about 1 Gyr, continuous minor mergers are necessary for steady magnetic field growth over several Gyrs \citep{2019MNRAS.486..623D}. Using the cryogenic X-ray microcalorimeter, soft X-ray spectrometer (SXS; \citealp{2016SPIE.9905E..0VK}), the \citet{2016Natur.535..117H} measured turbulent velocities in the Perseus cluster and found that the ICM was fairly quiescent, contradicting expectations. Synthesized observations \citep[e.g.][]{2018A&A...618A..39R} of the not-yet-launched Athena X-ray satellite show that the X-ray Integral Field Unit spectrometer will have both unprecedented spectral resolution (2.5 eV at 7 keV) and spatial resolution ($\sim$ few kpc) in measuring turbulence, even out to the cluster outskirts \citep[see][for a recent review]{2019SSRv..215...24S}. 

An ongoing project for the galaxy cluster science community is to gather sufficient data on a large number of ICM sources to allow for more thorough statistical studies of clusters. The Giant Metrewave Radio Telescope (GMRT) radio halo survey \citep{2008A&A...484..327V} and extended GMRT radio halo survey \citep{2013A&A...557A..99K} started this pursuit, providing observations of more than 60 galaxy clusters and finding radio halos in about $\sim 23 \%$ of clusters. The LOFAR Two Metre Sky Survey (LoTSS; \citealp{2017A&A...598A.104S}) is currently in progress to cover the entire northern hemisphere with very high sensitivity ($100~\mu$Jy~$/$~beam) at 140 MHz. Van Weeren et al., (in prep) present observations of more than 50 clusters covered in a 400 square degree region of LoTSS, listing all candidates for newly-discovered diffuse cluster emission. In the southern hemisphere, Johnston-Hollitt et al., (in prep) are using the GaLactic and Extra-galactic All-sky Murchison Widefield Array (MWA) survey (GLEAM; \citealp{2015PASA...32...25W}) to search for and identify candidate diffuse emission in 1,167 MCXC galaxy clusters (Meta-Catalogue of X-ray detected Clusters; \citealp{2011A&A...534A.109P}). Clusters selected from the South Pole Telescope (SPT) catalogue, which has detected over 1,000 clusters via the Sunyaev Ze'ldovich (SZ) effect since 2015 \citep{2015ApJS..216...27B, 2019ApJ...878...55B, 2019arXiv191004121B}, will be covered by the future surveys of the Square Kilometre Array \citep{2009IEEEP..97.1482D}. 

The Australian Square Kilometre Array Pathfinder (ASKAP; \citealp{2007PASA...24..174J, 2008ExA....22..151J}) is a newly commissioned radio telescope array consisting of 36 separate 12-m parabolic dish antennas operating at 700-1800 MHz in Western Australia. ASKAP stands apart from its predecessors due to its highly innovative Phased Array Feeds (PAFs) installed on each antenna. The PAFs are designed as a dual-polarisation chequerboard grid consisting of 188 element sensors that are cross-correlated to form 36 separate beams on the sky \citep{2014PASA...31...41H, 2016PASA...33...42M}. This technology gives the telescope a large field of view ideal for rapid survey imaging \citep{2009IEEEP..97.1507D}. The Evolutionary Map of the Universe survey (EMU; \citealp{2011PASA...28..215N}) will record the radio continuum over the whole Southern sky and up to +30 declination. Science goals of the EMU collaboration include tracing the evolution of galaxies and black holes, and further constraining cosmological parameters based on observations of large-scale structure. EMU, in conjunction with the Polarisation Sky Survey POSSUM \citep{2010AAS...21547013G}, will uncover more details on cosmic magnetism. Predictions have been made that the ASKAP EMU survey will detect more than 100 new radio halos \citep{2012A&A...548A.100C}, and recent work has shown that ASKAP has superb diffuse source sensitivity to emission on clusters scales (Hodgson et al., in prep). An EMU Pilot Survey was carried out in 2019 for a total of 10 fields and was made publicly accessible in the form of $\sim$~30 square degree mosaic images and calibrated visibilities in the \href{https://data.csiro.au/collections/#domain/casdaObservation/search/}{CSIRO ASKAP Science Data Archive} (CASDA; \citealp{2017ASPC..512...73C}).

In this paper we report on radio emission associated with two galaxy clusters that have been covered by ASKAP's early science observations outside of the EMU Pilot Survey. These galaxy clusters are high-mass, merging clusters that have been detected by the SPT through their SZ signal: SPT CLJ0553-3342 (also known as MACSJ0553.4-3342; hereafter MACSJ0553) and SPT CLJ0638-5358 (also known as Abell S0592; hereafter AS0592). Due to their high-mass both of these clusters are gravitational lenses, and have been imaged with the Hubble Space Telescope (HST) Advanced Camera for Surveys\footnote{MACSJ0553 was imaged in band $F435W$, $F606W$, and $F814W$, while AS0592 was only imaged in band $F606W$.} \citep[ACS;][]{1998SPIE.3356..234F}. The publicly available ASKAP images covering these clusters show clear signs of diffuse emission present in both intracluster media. See Table~\ref{tab1} for quantitative details on both clusters.

In the following subsections we outline prior scientific results found in the literature on both clusters. In Sect.~\ref{sec:methods} we describe the procedure we developed to carry out direction-dependent calibration and imaging on ASKAP data, as well as the data analysis methods we used to measure the properties of the diffuse emission from ASKAP in conjunction with supplementary data from other telescopes. Our findings are presented in Sect.~\ref{sec:results} and we end with a discussion and conclusion in Sects.~\ref{sec:discussion} and~\ref{sec:conclusions}. Throughout this paper we assume a $\Lambda$CDM cosmology with $H_{0} = 70$~km~s$^{-1}$~Mpc$^{-1}$, $\Omega_{m} = 0.3$, and $\Omega_{\Lambda} = 0.7$. All images are in the J2000 coordinate system.

\begin{table*}
 \centering
  \caption{Cluster properties from SPT and Planck catalogues and Chandra observations. See Sects.~\ref{macsintro} and \ref{abellintro} for references.} \label{table}
   \label{tab1}
  \begin{tabular}{c|c|c|c|c|c|c}
  \hline
Cluster & R.A., Dec. & SPT $M_{500}$ & Planck $M_{500}$ & $z$ & $L_{X}$ & T \\
 &  & [$10^{14}$ M$_{\odot}$] & [$10^{14}$ M$_{\odot}$] &  & [$10^{44}$~ergs sec$^{-1}$] & [keV] \\
\hline
\hline \\
AS0592 & 06h38m47.4s, $-53^{\circ}58''29.6'$ & $11.29^{+1.36}_{-1.10}$ & $6.83^{+0.34}_{-0.31}$ & 0.226 & $11.2 \pm 0.6$ & $9.45 \pm 0.95$\\ \\
MACSJ0553 & 05h53m24.3s, $-33^{\circ}42''43.4'$ & $11.33^{+1.37}_{-1.16}$ & $9.39^{+0.56}_{-0.58}$ & 0.412 & $10.3 \pm 0.3$ & $12.08 \pm 0.63$ \\ \\
\hline
\end{tabular}
\end{table*}

\subsection{MACSJ0553.4-3342}\label{macsintro}
MACSJ0553, also classified as SPT CLJ0553-3342, was first discovered in the Massive Cluster Survey (MACS; \citealp{2001ApJ...553..668E}). It is a massive, dynamically disturbed merging cluster, at a redshift of $z=0.407$  \citep{2008ApJ...682..821C}, and has been thoroughly researched at optical and X-ray wavelengths. \citet{2012MNRAS.420.2120M} preformed a joint X-ray-optical analysis and suggested that the two X-ray peaks visible in Chandra observations represent a binary head-on merger of two similar mass clusters with a merging axis in the plane of the sky. They also assessed that the merger evolutionary stage is likely after core passage. A further, more detailed study of this system was published in \citet{2017MNRAS.471.3305E} where the dynamics of dark and luminous matter were considered. There they combined HST and Chandra data and found that the merger axis is actually not in the plane of the sky but at a large inclination angle, and that the less massive Western component was fully stripped by ram-pressure as it passed through the more massive Eastern subcluster. 

Using a Chandra observation with longer exposure, \citet{2017MNRAS.472.2042P} measured and mapped discontinuities in surface brightness and temperature throughout the system and found evidence of two edges corresponding to one cold front and one shock on the Eastern side of the cluster. They report that the merger-driven cold front is behind the shock, following the morphology of other similarly merging clusters such as the \textit{Bullet} \citep{2002ApJ...567L..27M} and \textit{Toothbrush} \citep{2009A&A...506.1083V} clusters. They calculated the shock Mach number to be $1.33< \mathcal{M} < 1.72$. Both \citet{2017MNRAS.472.2042P} and \citet{2017MNRAS.471.3305E} confirmed the presence of this cold front and shock and used HST data to surmise that there are two subclusters of galaxies that make up this system, SC1 and SC2, which are separated by a projected distance of 650~kpc. More recently, \citet{2018MNRAS.476.5591B} used an edge detection filter and spectral analysis to search for shocks and cold fronts in a sample of 15 mass-selected clusters, including MACSJ0553, and they found that the cluster additionally hosts another cold front on the opposite, Western side of its ICM.

The most recent mass estimate of this cluster is $M_{500} = 11.33^{+1.37}_{-1.16} \times 10^{14}$ M$_{\odot}$ from the SPTpol Extended Cluster Survey catalog \citep{2019arXiv191004121B}. However, the mass estimated from the Planck satellite is lower: $M_{500} = 9.39^{+0.56}_{-0.58} \times 10^{14}$ M$_{\odot}$ \citep[PZS1;][]{2015A&A...581A..14P}. From the calculations of \citet{2017MNRAS.472.2042P},
MACSJ0553 is one of the hottest and most luminous galaxy clusters known ($T = 12.08 \pm 0.63$ keV and $L_{500,[0.1-2.4 keV]} = (10.2 \pm 0.3) \times 10^{44}$ erg s$^{-1}$).

\citet{2012MNRAS.426...40B} carried out a radio study of four massive clusters at redshifts $z>0.3$ with GMRT observations at 323 MHz and found an extended radio halo in MACSJ0553. While two other clusters in their sample showed clear double radio relics, there were no such radio shock structures observed in MACSJ0553. They elaborated upon the unexpected absence of radio relics in this cluster, predicting that the merger axis may not be in the plane of the sky (which was later confirmed by \citealp{2017MNRAS.471.3305E}) and therefore it may be difficult to see radio shocks due to the combination of projection effects and the brightness of the halo. 

\subsection{Abell S0592}\label{abellintro}
AS0592, also classified as SPT CLJ0638-5358 and RXC J0638.7-5358 (from the REXCESS survey; \citealp{2007A&A...469..363B}), is a massive cluster with a recent mass estimate of $M_{500} = 11.29^{+1.36}_{-1.10} \times 10^{14}$ M$_{\odot}$ from the SPT-SZ 2500~deg$^{2}$ catalogue \citep{Bocquet_2019}. A previous estimate of the cluster mass, from the Planck catalogue, is much lower: $M_{500} = 6.83^{+0.34}_{-0.31} \times 10^{14}$ M$_{\odot}$ \citep[PSZ2;][]{2016A&A...594A..27P}. This cluster was covered by the ROSAT REFLEX Survey, from which the redshift was estimated to be $z=0.221$ \citep{1999ApJ...514..148D}. An update to the redshift was provided by \citet{2011A&A...534A.109P} where they found $z=0.226$. An X-ray analysis for this cluster was published in \citet{2009PhDT........18M} where the following values were measured from a ACIS-I VFAINT 19.9~ks follow-up Chandra observation: $kT = 9.5 \pm 1.0 $ keV, L$_{X} = (11.2 \pm 0.6)\times 10^{44}$~ergs~sec$^{-1}$, and  $M_{500} = (10.3 \pm 1.4) \times 10^{14}$ M$_{\odot}$. 

Although it has a lower temperature and a slightly lower mass (according the SPT measurement), AS0592 is more X-ray luminous than MACSJ0553. In a proposal for deep XMM-Newton observations, \citet{2009xmm..prop...26H} suggested that archival Chandra and HST data reveal that AS0592 is undergoing a major merger. In a study of the cool-core state of Planck-selected clusters, \citet{2017MNRAS.468.1917R} used X-ray observations of AS0592 to measure the concentration parameter as defined by \citet{2008A&A...483...35S}. They found that the cluster has a significant surface brightness peak but an overall disturbed X-ray morphology, and they list it as a ``disturbed cool-core.'' \citet{2018MNRAS.476.5591B} combined four separate Chandra observations of AS0592 (for a total net exposure time of 98 ks) and noted the presence of two low-temperature, low-entropy cool-cores surrounded by a hot, disturbed ICM. They also claim that there is a shock along the Southwest edge of the ICM with a Mach number $1.61 < \mathcal{M} < 1.72$.

So far, there have been no in-depth radio studies published for this cluster, but since the X-ray studies prove it to be dynamically disturbed and hosting a shock, it is a good candidate for exhibiting diffuse intracluster emission. 

\section{METHODS}\label{sec:methods}

\subsection{ASKAP observations and pre-processing}

Data for both clusters was obtained from \href{https://data.csiro.au/collections/#domain/casdaObservation/search/}{CASDA}\citep{2017ASPC..512...73C}, and come from ASKAP's early science observations Scheduling Block (SB) 8275 and SB9596 for AS0592 and MACSJ0553, respectively. SB8275 is part of the ASKAP Early Science Broadband Survey (Project: AS034, PI: Lisa Harvey-Smith) and SB9596 is part of the ASKAP Pilot Survey for Gravitational Wave Counterparts (Project: AS111, PI: Tara Murphy). For each SB, there are 36 measurements sets corresponding to 36 separate beams (covering $\sim$6 deg$^{2}$ in the sky) observed by each of the 36 antennas. MACSJ0553 and Abell S0592 fall within the relative centre of a single beam pointing for their respective survey observations\footnote{MACSJ0553 falls within the relative centre of Beam 14 of SB9596 and AS0592 falls within the relative centre of Beam 22 of SB8275.}. Each beam measurement set in a Scheduling Block has undergone direction-\textit{independent} calibration as part of the ASKAP processing pipeline (ASKAPsoft; \citealp{2019ascl.soft12003G}). For each SB, CASDA also provides a mosaicked image for an overall field of view of $\sim$36 deg$^{2}$.

In the following paragraph, we provide an unofficial summary of the ASKAPsoft processing pipeline for pre-calibration and imaging of ASKAP's early science observations, however, we also refer the reader to \href{https://www.atnf.csiro.au/computing/software/askapsoft/sdp/docs/current/pipelines/introduction.html}{CSIRO's ASKAPsoft documentation} for further details. For direction-\textit{independent} calibration, ASKAPsoft performs bandpass calibration using the standard calibrator PKS B1934-638, which is observed for five minutes in each beam before or after the science target. After bandpass calibration the 36 beams are imaged and deconvolved independently. For every 1 MHz of data, a cycle of self-calibration and gains calibration is also applied. The primary beams are modeled as circular Gaussians with a size taken from holography observations of the true beam\footnote{See the memo: \href{https://www.atnf.csiro.au/projects/askap/aces_memo011.pdf}{Holographic Measurement of ASKAP Primary Beams.}}, and an average value is used for all 36 with an error of 10\%. Image weighting is implemented through a Wiener pre-conditioning method which has a setting equivalent to Brigg's robustness. As a final step, the 36 pre-calibrated beam images are stitched together to form a linear mosaic in the image plane. CSIRO has provided these pre-calibrated data sets as well as the mosaicked image on CASDA, but state that early science observations have not been validated.\\

\textbf{Official statement on validation from CASDA:} 
\begin{quote}
    While [these data] may be useful for science purposes the primary goal of the release is to seek feedback on data quality from the science community. Although the image approaches expected noise levels in most areas, a common feature of early ASKAP data has been some low-level artefacts (at ~1\%) very close to bright sources (a few hundred mJy and above). These artefacts generally appear as radial stripes and rapidly fade away from the source. The origin of these artefacts is yet to be determined, but they do not seem to emulate any known instrumental effect taken into account in commonly used wideband/widefield calibration and imaging software packages (such as ASKAPsoft, casa etc.)[...] While the commissioning team investigates the cause of these issues in detail, we do not wish to hold the release of otherwise good quality data. Hence these test observations are being released, but with the validation flag set to ``uncertain'' -- meaning user discretion is necessary in the interpretation for science purposes [...] -- Emil Lenc / Aidan Hotan 
\end{quote}

\subsection{Additional direction-dependent processing}

To test whether we could improve the image quality of early science ASKAP data, we carried out direction-dependent calibration and imaging on individual pre-calibrated measurements sets taken from CASDA. We used third-generation calibration (3GC) software {\sc DDFacet} ({\sc DDF}) \citep{2018A&A...611A..87T} and {\sc killMS} ({\sc kMS}) \citep{2014A&A...566A.127T, 2015MNRAS.449.2668S} and packages therein, to image, compute, and apply solutions for direction-dependent errors. This software is currently being utilised as part of the official processing pipeline for LoTSS \citep[see][for details]{2019A&A...622A...1S}.

To make comparisons with the ASKAPsoft mosaics, we calibrated and imaged all 36 beams of both SB8275 and SB9596 independently and created a mosaic in the image plane. Since our target clusters fell within or near the centre of a single beam observation, we only used a single beam image, rather than the full mosaic, to produce science images of MACSJ0553 and Abell S0592. We note that the pre-calibrated measurement set covering MACSJ0553 (Beam 14 of SB9596) contained flags for antenna AK33 and the pre-calibrated measurement set covering AS0592 (Beam 22 of SB8275) contained flags for antenna AK03. We also note that the central frequencies of these measurement sets are different: 943~MHz for MACSJ0553 and 1013~MHz for AS0592. To test the ability of {\sc kMS} and {\sc DDF} in reducing artefacts around bright sources, we also calibrated and imaged individual beam pointings of other ASKAP observations (including SB9325 and SB9351). 

Our techniques for implementing direction-dependent calibration on these early science observations have been modified and streamlined into a processing pipeline written in python, which accepts any early science ASKAP measurement sets and produces a final mosaicked image after correcting for directional effects in each of the 36 beams. Plans to expand the pipeline and make it open source are underway. We refer the reader to \citet{2015MNRAS.449.2668S} and  \citet{2018A&A...611A..87T} for the full details of the mathematical functions implemented by the {\sc kMS} and {\sc DDF} software, but here we briefly explain the problem and benefit of solving for directional errors/effects. 

\begin{figure*}
\centering
\includegraphics[width=0.49\textwidth]{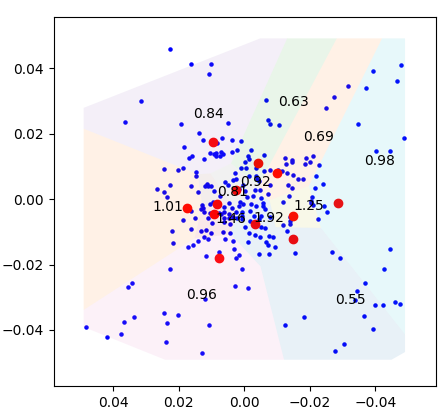}
\includegraphics[width=0.49\textwidth]{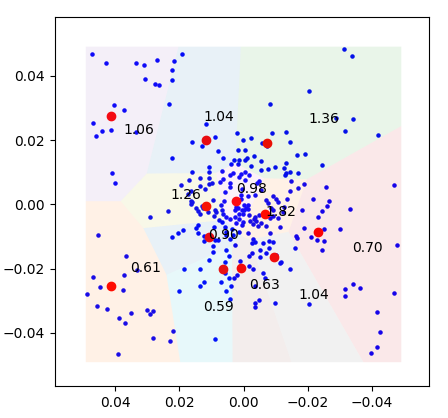}
\caption{ The directional facet scheme for {\sc killMS} calibration of the beams containing AS0592 (left) and MACS0553 (right). Sources (blue dots) are grouped into facets using a Voronoi tessellation algorithm, and calibrators (red dots with flux listed in Jy) are used to compute solutions for each facet. Axes are in radians from the centre.  \label{facets}}
\end{figure*}

In radio observations the true radio signal from the sky is perturbed by both direction-independent (DI) and direction-dependent (DD) effects. The electronics of the instrumentation (as the signal goes from the amplifier to the voltage multiplier, for example) and small discrepancies in station clocks account for the DI effects. More complex are the DD effects, which include station beams, the ionosphere, and Faraday rotation. The Radio Interferometry Measurement Equation (RIME; \citealp{1996AAS..117..137H}) models these effects as linear transformations, also called Jones matrices (\textbf{F} for DI effects and \textbf{G} for DD effects, as expressed below), so that they can be solved for, and the visibility function, \textbf{V}$_{(p,q),t\nu}$ between antennas $p$ and $q$ at time $t$ and frequency $\nu$, can be transformed into the true underlying source coherency matrix \textbf{X}$_s$. From the \textit{sky term}, \textbf{X}$_s$, a spatial intensity image of the radio signal in the direction $\boldsymbol{s}$ $= \left[{\rm l, m, n} = \sqrt{1 - l^2 - m^2}\right]^T$ can be produced \citep[e.g.][]{2018A&A...611A..87T}: 
 
 \begin{equation}
 \label{rime}
\boldsymbol{\rm V}_{(p,q),t\nu} = \boldsymbol{\rm F}_{pt\nu} \left[\int_s (\boldsymbol{\rm G}_{pst\nu}\boldsymbol{\rm X}_{s}\boldsymbol{\rm G}^{H}_{qst\nu})k^{s}_{(p,q),t\nu} d\boldsymbol{s}\right]\boldsymbol{\rm F}^{H}_{qt\nu},
\end{equation}
where $ k^{s}_{(p,q),t\nu}$ describes the effect of the array geometry and correlator on the observed phase shift of a coherent plane wave between antennas $p$ and $q$:
\begin{equation}\label{eq1}
    k^{s}_{(p,q),t\nu} = {\rm exp}\left( -2i\pi \frac{\nu}{c}\left(\boldsymbol{b}^T_{(p,q)t}(\boldsymbol{s} - \boldsymbol{s_0})\right)\right),
\end{equation}
\begin{align}
    \boldsymbol{b}_{(p,q)t} &= \begin{bmatrix}
           u_{pq,t} \\
           v_{pq,t} \\
           w_{pq,t}
         \end{bmatrix}.
  \end{align}
  
The integral in Eq.~\ref{eq1} requires that the Jones Matrices \textbf{G} must be solved for all infinitesimal directions in the sky. To approximate this integral, the \textit{faceting} technique is used, where the observational field of view is split up piece-wise into several smaller directions represented by facets (polygonal regions). The {\sc killMS} software calculates gains solutions on each smaller direction in parallel. The {\sc DDFacet} imager can then image the visibilities while applying the solutions to each facet region in the sky. 
The below list summarizes our full processing steps on the ASKAP early science data to produce science quality images of the galaxy  clusters:
\begin{itemize}
    \item Before running additional processing we shifted the pointing centre of the observations by the beam offset. Offsets are retrieved from the ``FEED'' table of the measurement set and the beam position is updated for each field. Here we note that for all pre-calibrated measurement sets pulled from CASDA, the Stokes I value of ASKAP visibility data is $1/2$ of the true intensity value. 
    
    \item
    Using {\tt DDF.py}, we deconvolved the pre-calibrated measurement sets covering MACSJ0553 and AS0592 with the auto-mask option enabled. We then used {\tt MakeMask.py} to produce a mask for all islands above $10\sigma$. We deconvolved once more using the generated mask to produce a DI image. 
    
    \item In the next step, we used {\tt MakeCatalog.py} on the DI image to identify sources that will be utilised as directional calibrators. {\tt ClusterCat.py} then tessellates the field of view into seven facet directions, each containing sufficient flux for DD calibration. See Fig.~\ref{facets} for a plot of the faceting schemes used for the fields containing AS0592 and MACS0553. The faceting area covers $\sim~25$ square degrees.
    
    \item Using the model from the last DI image and the catalogue and faceting map from the last step, we ran {\tt kMS.py} to compute directional amplitude and phase solutions. The ``KAFCA scalar'' solver mode was selected and the solution intervals for time and frequency were set to five minutes and 30~MHz, respectively. These solutions were saved within a table of the measurement set. 
    
    \item We then deconvolved using {\tt DDF.py} while applying the solutions from {\sc kMS} to make a DD image. To ensure that diffuse emission was properly deconvolved, we made a new mask of all islands above $4\sigma$ including a region around the entire cluster. The imaging parameters for the DD image include a selection of baselines greater than 60 meters, a cell size of 2 arcsec, an image size of [12000, 12000] pixels, a Brigg's robust parameter of -1.5, and a restoring beam of 11 arcsec, close to the native resolution. These parameters gave results that best matched the ASKAPsoft mosaic image.
    
    \item Our final step included applying a primary beam correction, assuming a Gaussian with a FWHM of $1.09\lambda / D$, and multiplying the Stoke's I visibilities by two. 
    
\end{itemize}
Further steps were taken to properly subtract emission from AGN within and near the clusters, so that the diffuse emission could be accurately measured. Subtraction methods are described in the following subsection. Additional images were made after subtraction, and those imaging parameters are described in Sect.~\ref{sec:results} for each cluster. No astrometric offset was perceptible when comparing source positions in optical and radio maps at varying frequencies. To determine the error in the flux scale of our final ASKAP images, we created a model catalogue from a combined GLEAM extragalactic catalogue (GLEAM EGC; \citealp{2017MNRAS.464.1146H}) and Sydney University Molonglo Sky Survey (SUMSS; \citealp{2003MNRAS.342.1117M}) catalogue, and extrapolated flux densities to ASKAP frequencies. For the field containing MACSJ0553 we found that the error on the flux was $\sim~10\%$ and for the field containing AS05922 we found the error to be $\sim~5\%$. 

\subsection{Subtraction of discrete sources}\label{sub}
To accurately measure diffuse emission associated with the ICM in these clusters, we performed a subtraction of flux from sufficiently bright, compact AGN in the cluster environments. For point sources, measuring and modeling the peak flux across several sub-band images was sufficient. For more extended sources, several methods were attempted to accurately model the emission and remove the corresponding visibilities without introducing negative artefacts. When modeling sources to be subtracted we used images with minimum baseline cuts to capture emission on scales $< 250$~kpc. Based on the redshifts of the galaxy clusters, this corresponded to $>4500~\lambda$ for MACSJ0553 and $>3000~\lambda$ for AS0592. 

For MACSJ0553, prior to DD calibration and imaging, we created six sub-band images (bandwidth of 48~MHz each) with {\tt WSClean} \citep{2014MNRAS.444..606O} and measured the peak flux density of point source AGN which appeared to eclipse diffuse emission. We then used {\tt Subtrmodel} \citep{2014MNRAS.444..606O} with a model file listing the measurements of the sources at the six different sub-band frequencies. {\tt Subtrmodel} performs a calculation of the spectral index of each source so that the flux can be removed from the visibilities across the full band, and then subtracts the modeled visibilities from the data column. 
Once sources were subtracted, we then preformed DD calibration on the modified measurement set and used the final image to measure the flux density of diffuse cluster-scale emission. 

For Abell S0592, the AGN embedded in diffuse emission were much more extended, and could not be modeled as point sources. These bright AGN were also producing slight artefacts, so we did not wish to subtract them prior to DD calibration. Instead, we carried out DD calibration, and then used {\tt DDF.py} and {\tt MakeMask.py} to create a compact-emission DD image with a customized mask covering the full extended emission from these radio galaxies. Using the Predict option of {\tt DDF.py}, the CLEAN components from the compact image were placed into a model column. We then manually subtracted this model column from the data column and re-imaged again while applying the {\sc kMS} solutions.  

\begin{figure*}
\begin{subfigure}{.5\textwidth}
  \centering
  \includegraphics[width=0.8\linewidth]{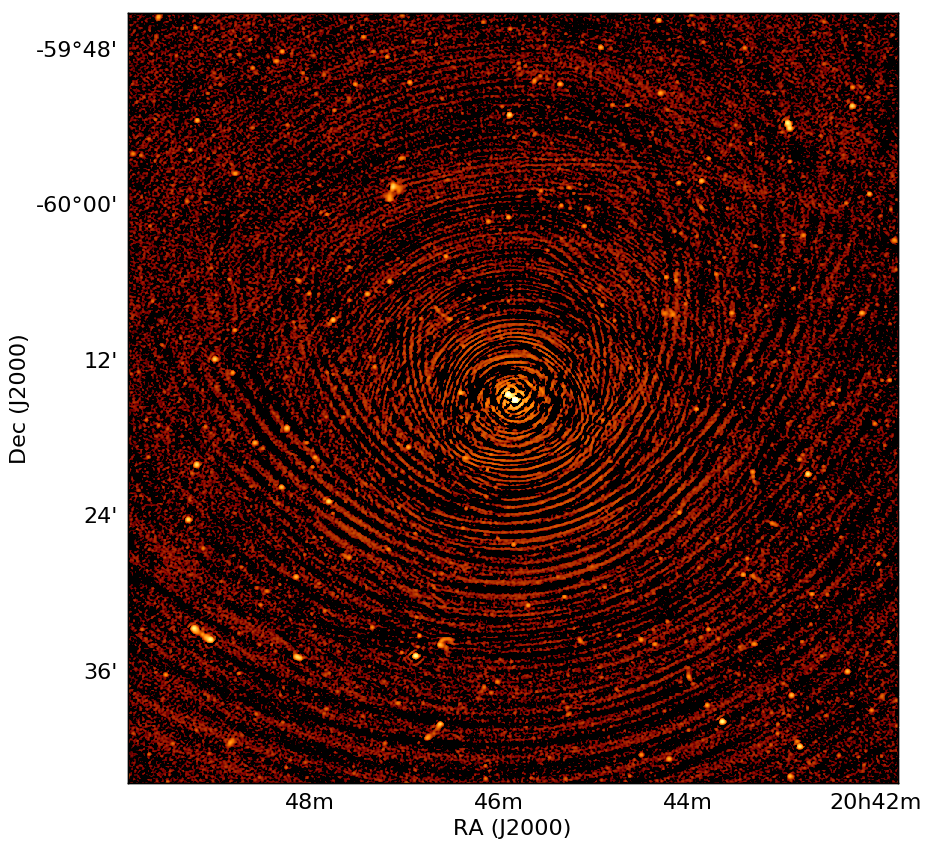}  
  \caption{ASKAPsoft image of PKS 2041-60.}
  \label{DDa}
\end{subfigure}
\begin{subfigure}{.5\textwidth}
  \centering
  \includegraphics[width=0.8\linewidth]{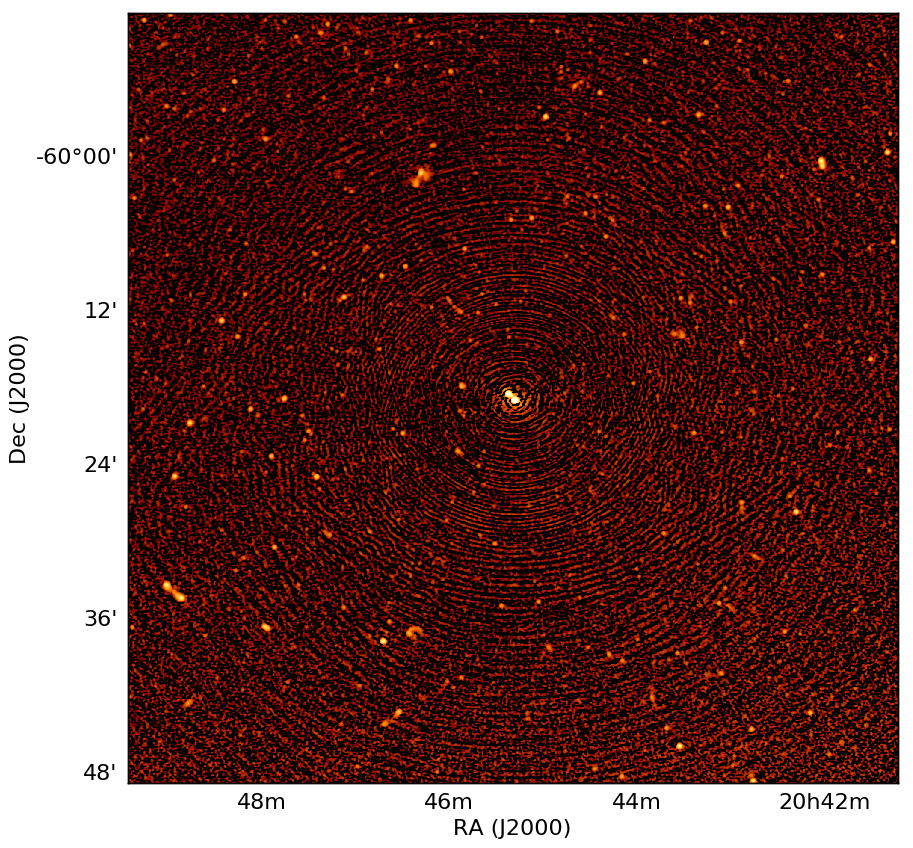}  
  \caption{{\sc DDF} image of PKS 2041-60.}
  \label{DDb}
\end{subfigure}


\begin{subfigure}{.5\textwidth}
  \centering
  \includegraphics[width=0.8\linewidth]{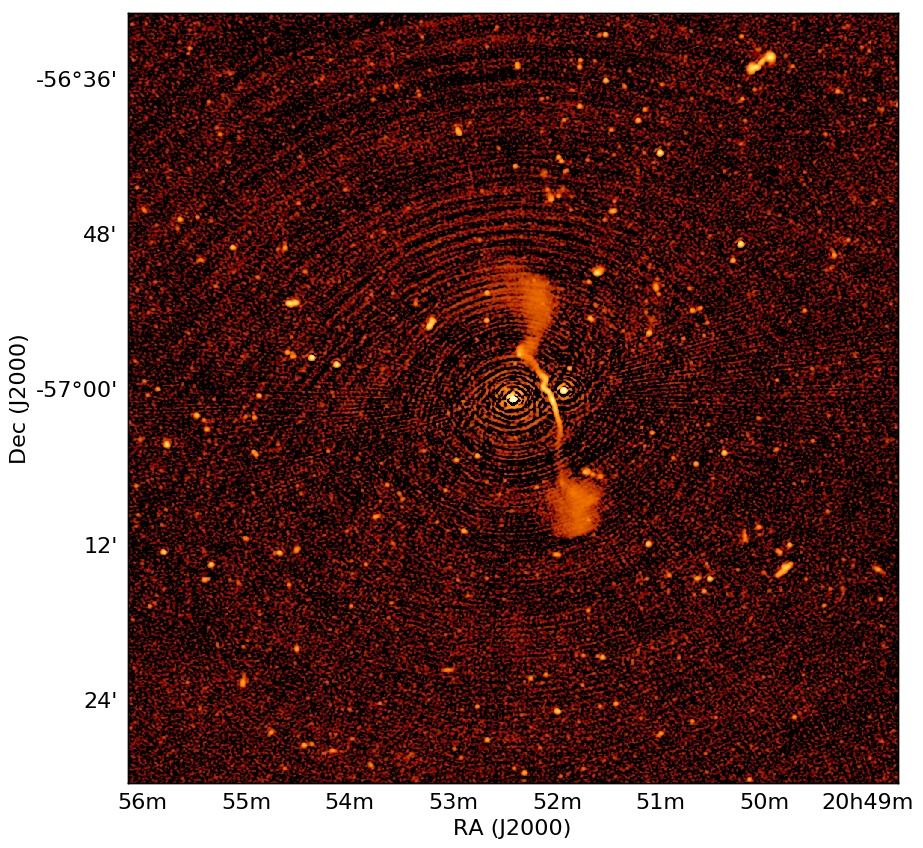}  
  \caption{ASKAPsoft image of IC 5063.}
  \label{DDc}
\end{subfigure}
\begin{subfigure}{.5\textwidth}
  \centering
  \includegraphics[width=0.8\linewidth]{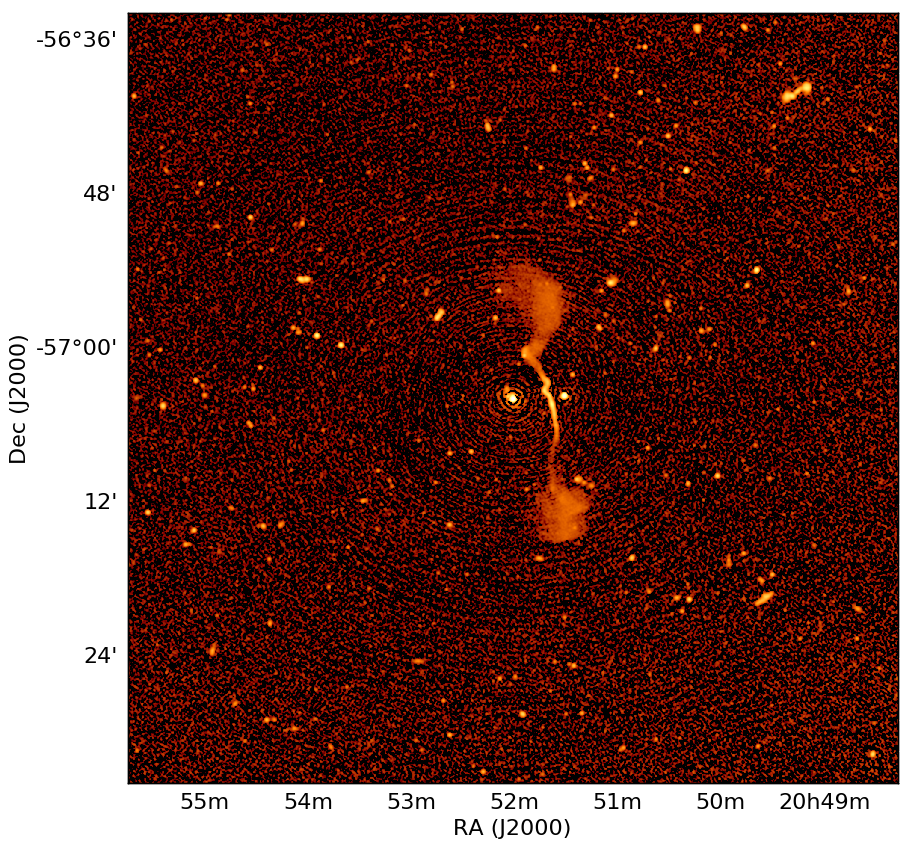}  
  \caption{{\sc DDF} image of IC 5063.}
  \label{DDd}
\end{subfigure}  
  

\begin{subfigure}{.5\textwidth}
  \centering
  \includegraphics[width=0.8\linewidth]{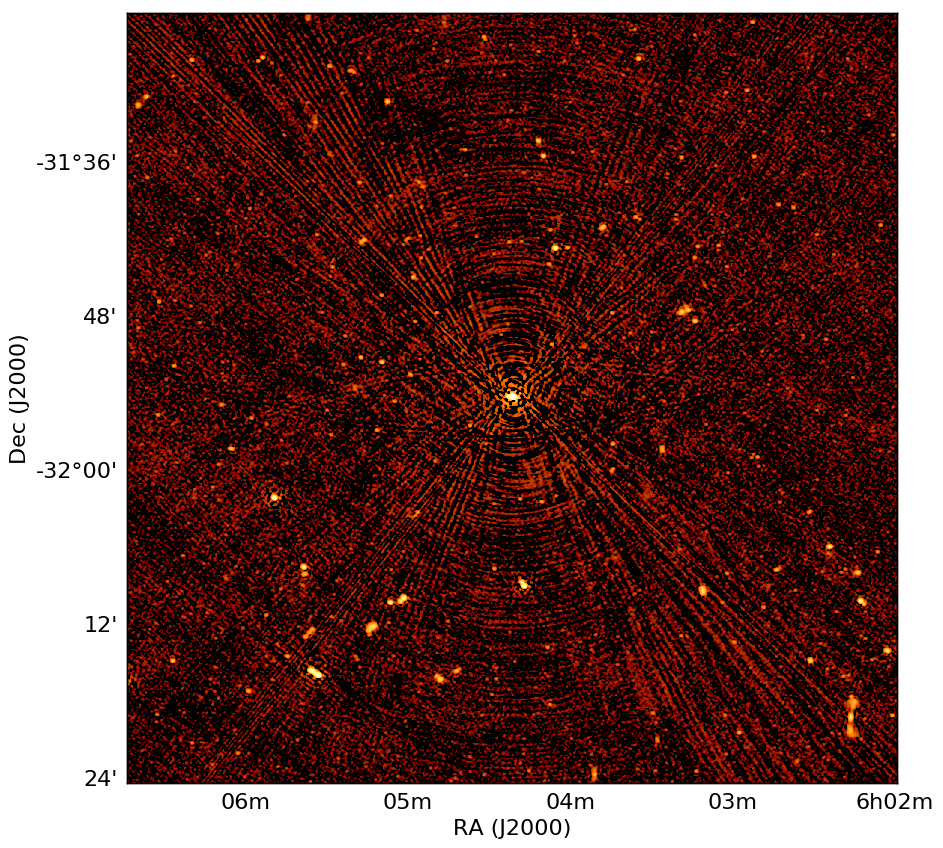}  
  \caption{ASKAPsoft image of QSO B0602-319.}
  \label{DDe}
\end{subfigure}
\begin{subfigure}{.5\textwidth}
  \centering
  \includegraphics[width=0.8\linewidth]{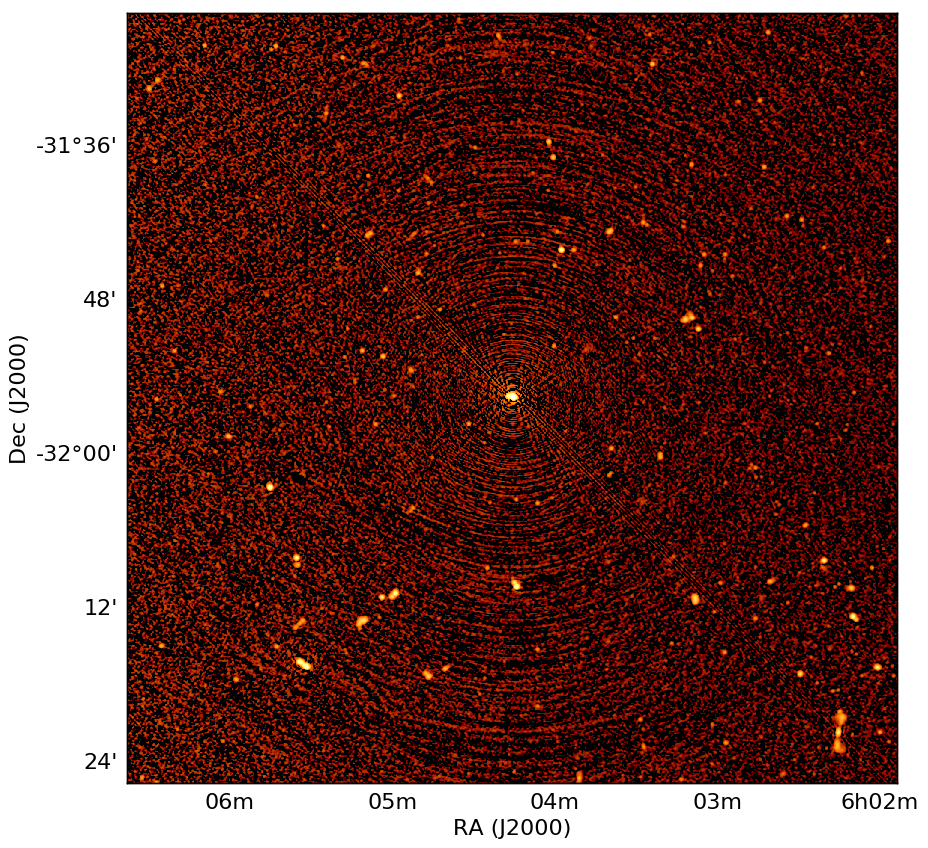}  
  \caption{{\sc DDF} image of QSO B0602-319.}
  \label{DDf}
\end{subfigure}

\caption{Bright sources before and after DD calibration. CSIRO's ASKAPsoft mosaic images, centred on bright radio sources producing prominent artefacts, are compared to our images after processing the same data with {\sc killMS} and {\sc DDFacet}. Images (a) and (b) come from the ASKAP pilot observation SB9325. Images (c) and (d) come from the ASKAP pilot observation SB9351. Images (e) and (f) come from the ASKAP pilot observation SB9596. Radial and stripe artefacts are greatly reduced after DD calibration. The colour scale is the same on all images, with a minimum value of $1~\mu$Jy and a maximum value of $1$~Jy.}
\label{DD}
\end{figure*}

\subsection{Supplementary data}

\subsubsection{X-ray}
We re-processed archival Chandra data (Obs IDs: 5813 and 16598 for MACSJ0553 and AS0592, respectively) for both clusters to make our own X-ray images. The Chandra data were reduced with {\tt CIAO} v4.10 with CALDB version 4.7.9. Periods of high background were detected with \textit{lc clean} using the S3 chip and the energy band 2.5-7~keV, and they were removed from the data. Since a thorough X-ray analysis of MACSJ0553 has already been published in \citet{2017MNRAS.472.2042P}, we found no reason to redo these calculations for the shock, and used their values in our anaylsis. The X-ray analysis for Abell S0592 is less thorough in the available literature, so we determined whether they system is dynamically disturbed and confirmed the discontinuity found by \citet{2018MNRAS.476.5591B}.

\subsubsection{Radio}
We re-processed and re-imaged the GMRT 323 MHz observations of MACSJ0553 published by \citet{2012MNRAS.426...40B} using the SPAM pipeline \citep[see][for details]{2017A&A...598A..78I}. We also imaged our SPAM calibrated data with CASA (Common Astronomy Software Applications; \citealp{2007ASPC..376..127M}) tools {\tt TCLEAN} deconvolver mode mtmfs with a $uv$ range selection of $>4500~\lambda$ to image compact emission only. Using CASA tools {\tt ft} and {\tt uvsub} the models from the compact image were placed in a model column of the measurement set and subtracted from the data column. We then produced a compact-source-subtracted image using {\tt TCLEAN} with a $uv$ range of $>100~\lambda$ and uniform weighting to allow for a comparison of the diffuse emission detected by ASKAP.

To identify potential point sources in the cluster, we obtained a S-band A-configuration VLA observation of MACSJ0553 with 196 minutes of integration time on the target field (Project Code: 17B-367), which we processed through the CASA VLA pipeline available in CASA 5.2.2. We imaged the processed data with CASA {\tt TCLEAN} deonvolver mtmfs.  

Abell~S0592 was observed with the Australia Telescope Compact Array \citep[ATCA;][]{fbw92} with the Compact Array Broadband Backend \citep[CABB;][]{cabb} for $\sim250$~minutes in the 16-cm band (Project Code: C2837). Data were accessed through the Australia Telescope Online Archive\footnote{\url{https://atoa.atnf.csiro.au/}} and initial bandpass and gain calibration was performed using the \texttt{miriad} software suite \citep{stw95}. The observations were taken in the EW352 and 750D array configurations, which each have a minimum baseline of 31~m, corresponding to angular scales of $\sim16$~arcmin at 2.1~GHz. Bandpass and absolute flux calibration was performed using the standard calibrator for ATCA cm observations, PKS~B1934$-$638, and the phase calibrator for the observation was PKS~0647$-$475. The data went through RFI flagging, and the original 2-GHz bandwidth was reduced to $\sim 1.8$~GHz.  

Due to the gap in the $uv$ coverage between the inner and outer baselines, we employed a $uv$ range selection of $< 10$~k$\lambda$ to ensure a well-behaved point-spread function. We used \texttt{CASA} and \texttt{WSClean} \citep{2014MNRAS.444..606O} to perform two rounds of phase-only self-calibration followed by a round of phase and amplitude self-calibration. During imaging the data were split into eight sub-bands of $\Delta\nu = 227.5$~MHz, though CLEANing was done by peak-finding on a full-bandwidth image to ensure faint point sources were deconvolved.

\section{RESULTS}\label{sec:results}

\subsection{Direction-dependent calibration}
 Some of the ASKAP early science and Pilot Survey images available on CASDA are affected by prominent artefacts from bright sources within the field. These artefacts are present after the data has been processed through the ASKAPsoft pipeline, which includes direction-\textit{independent} calibration only. To showcase the results of utilising the directional calibration and imaging software {\sc killMS} and {\sc DDFacet}, we provide a side-by-side comparison of images before and after DD calibration centred on three exemplary bright sources. Beam 22 of SB9325, hosting the bright source PKS 2041-60, beam 10 of SB9351, hosting the bright source IC 5063, and beam 28 of SB9596, hosting the bright quasar QSO B0602-319, all show radial artefacts (ring-type and/or stripe-type) in the ASKAPsoft image that increase the noise in a region of $\sim~15$ arcmin around those sources. In Fig.~\ref{DD} the ASKAPsoft images are compared to our {\sc DDF} images after applying DD calibration. There is a noticeable reduction of artefacts in the {\sc DDF} images. By checking the last DI residual and the first DD `dirty' image during our processing steps, we confirm that the improvement is due to applying the solutions from {\sc kMS} rather than from deeper deconvolution alone. We find that the root mean square (rms) noise, on average, is reduced from $\sim~40~\mu$Jy to $\sim~20~\mu$Jy, within 10~arcmin of the bright source.

\begin{figure*}
\centering
\includegraphics[width=0.49\textwidth]{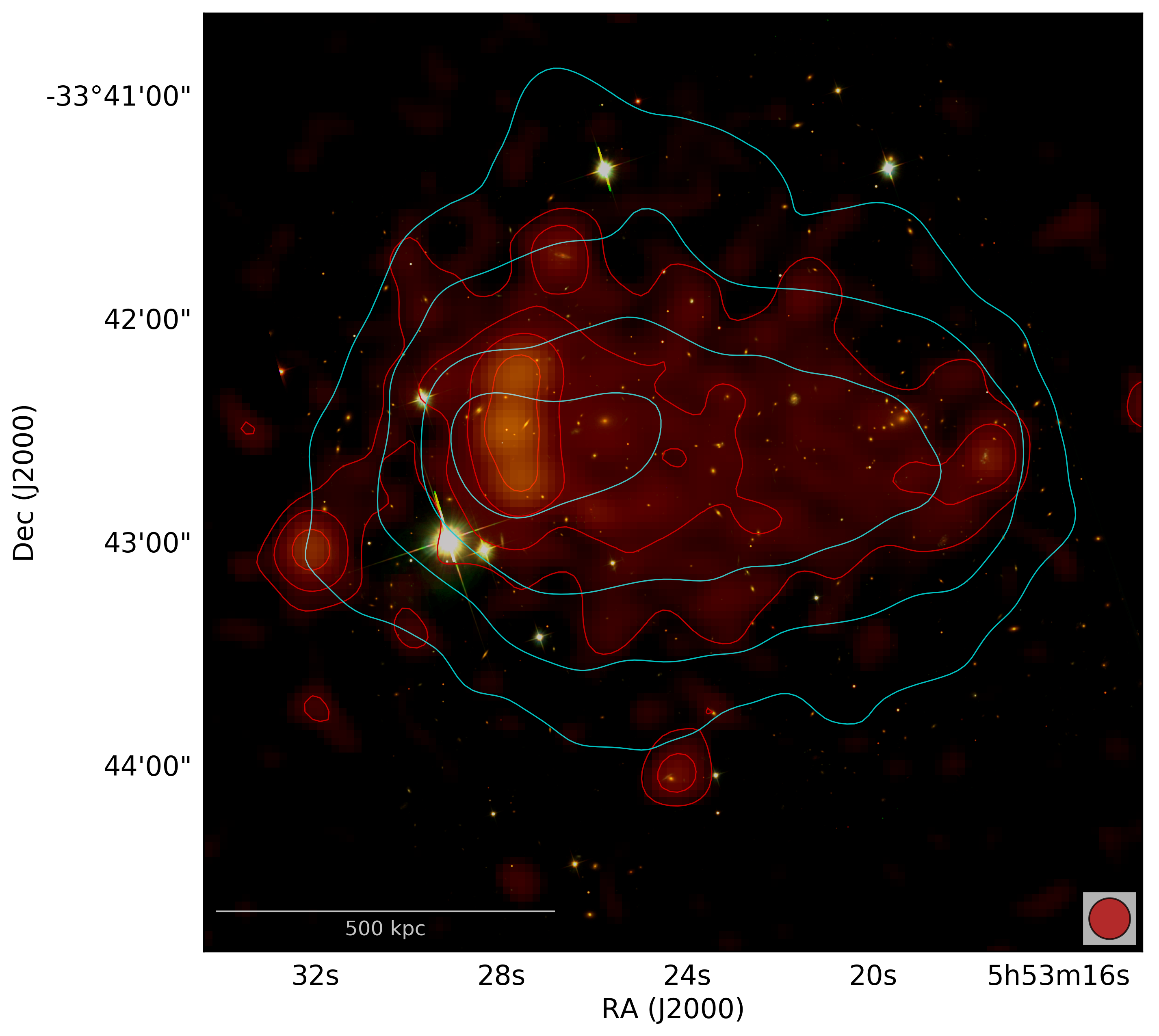}
\includegraphics[width=0.49\textwidth]{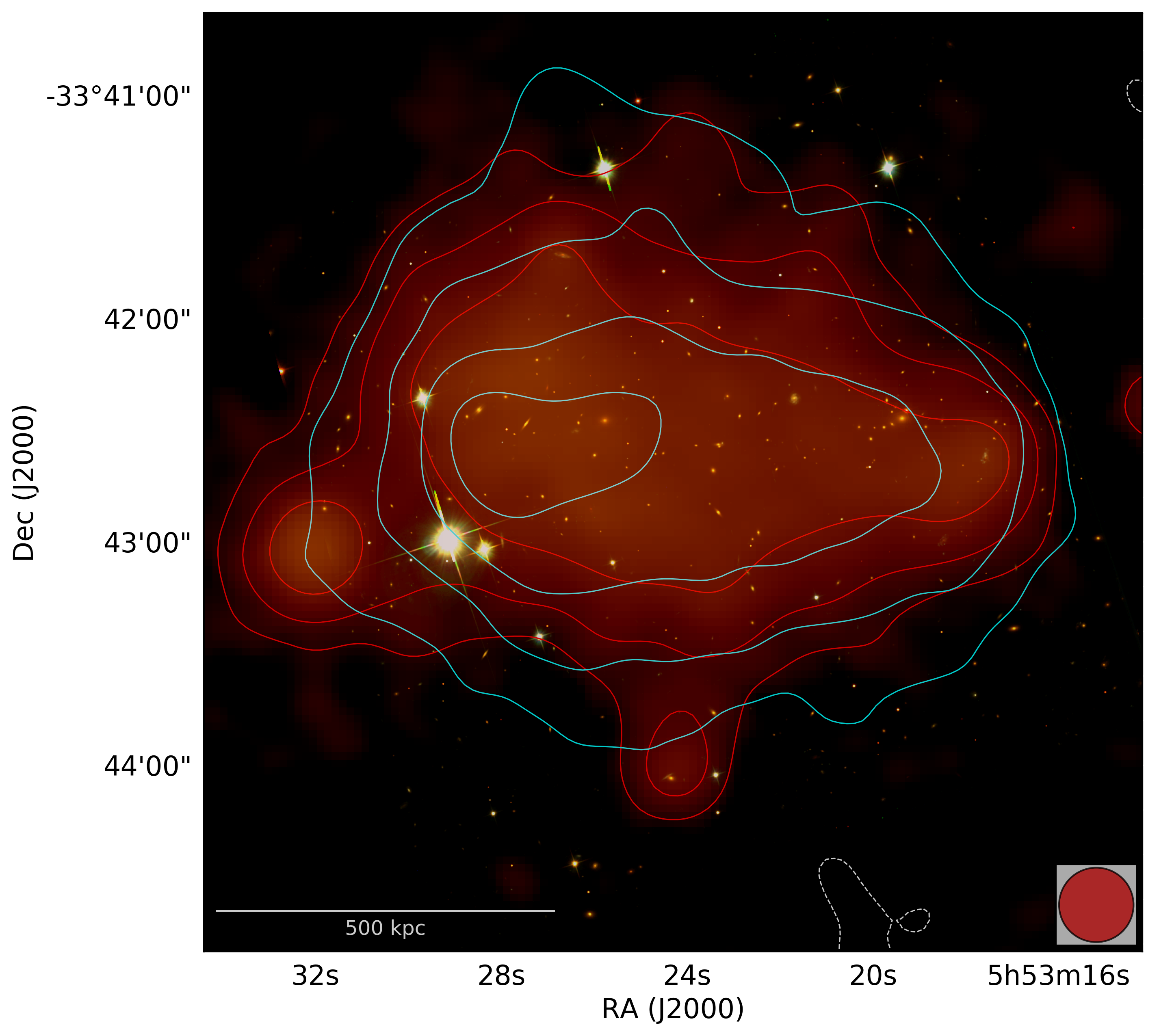}
\caption{HST {\it i,r,g} image of MACSJ0553 with  943~MHz ASKAP radio emission and Chandra X-ray emission overlaid as contours. ASKAP emission is shown by red contours at levels $[3, 6, 12, 24]~\times~\sigma$. Smoothed Chandra X-ray contours are in cyan. Right: Our ASKAP image made with {\sc DDF} after DD calibration ($\sigma = 20~\mu$Jy beam$^{-1}$, restoring beam 11 arcsec $\times$ 11 arcsec). Left: Our ASKAP image made with {\sc DDF} after point soucre subtraction and DD calibration ($\sigma = 25~\mu$Jy beam$^{-1}$, restoring beam 20 arcsec $\times$ 20 arcsec). The red colour of the ASKAP emission is included for visualization only. See text for imaging parameters.  \label{MACSJ0553opt}}
\end{figure*}

\begin{figure*}
\centering
\includegraphics[width=0.49\textwidth]{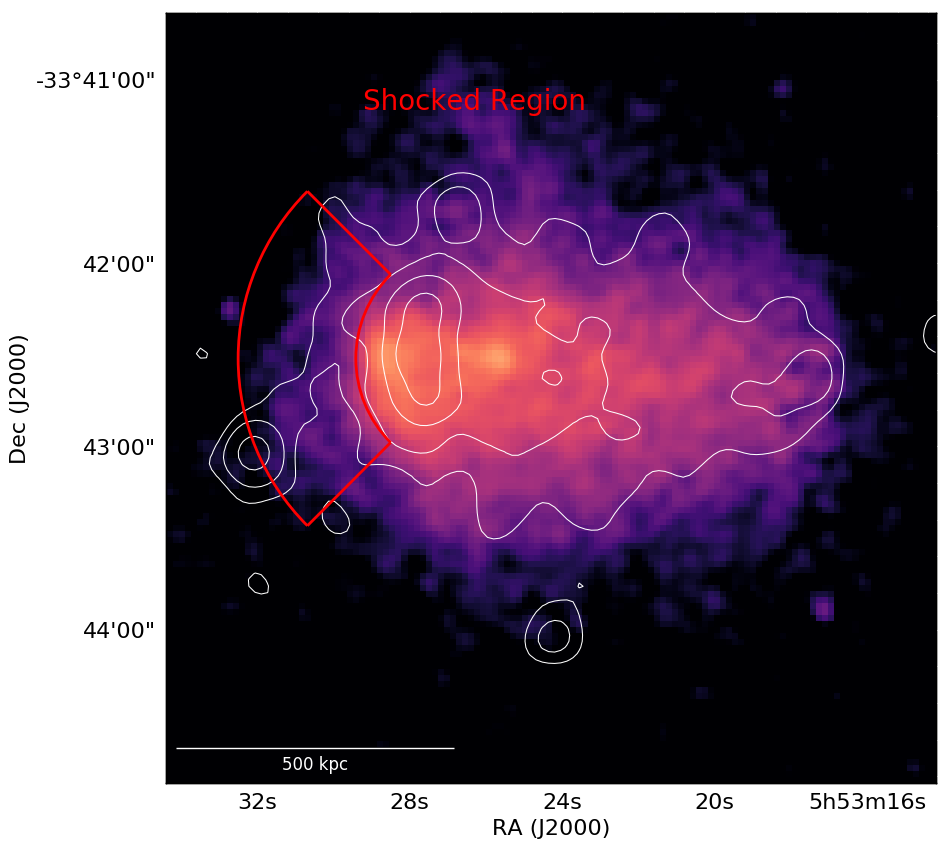}
\includegraphics[width=0.49\textwidth]{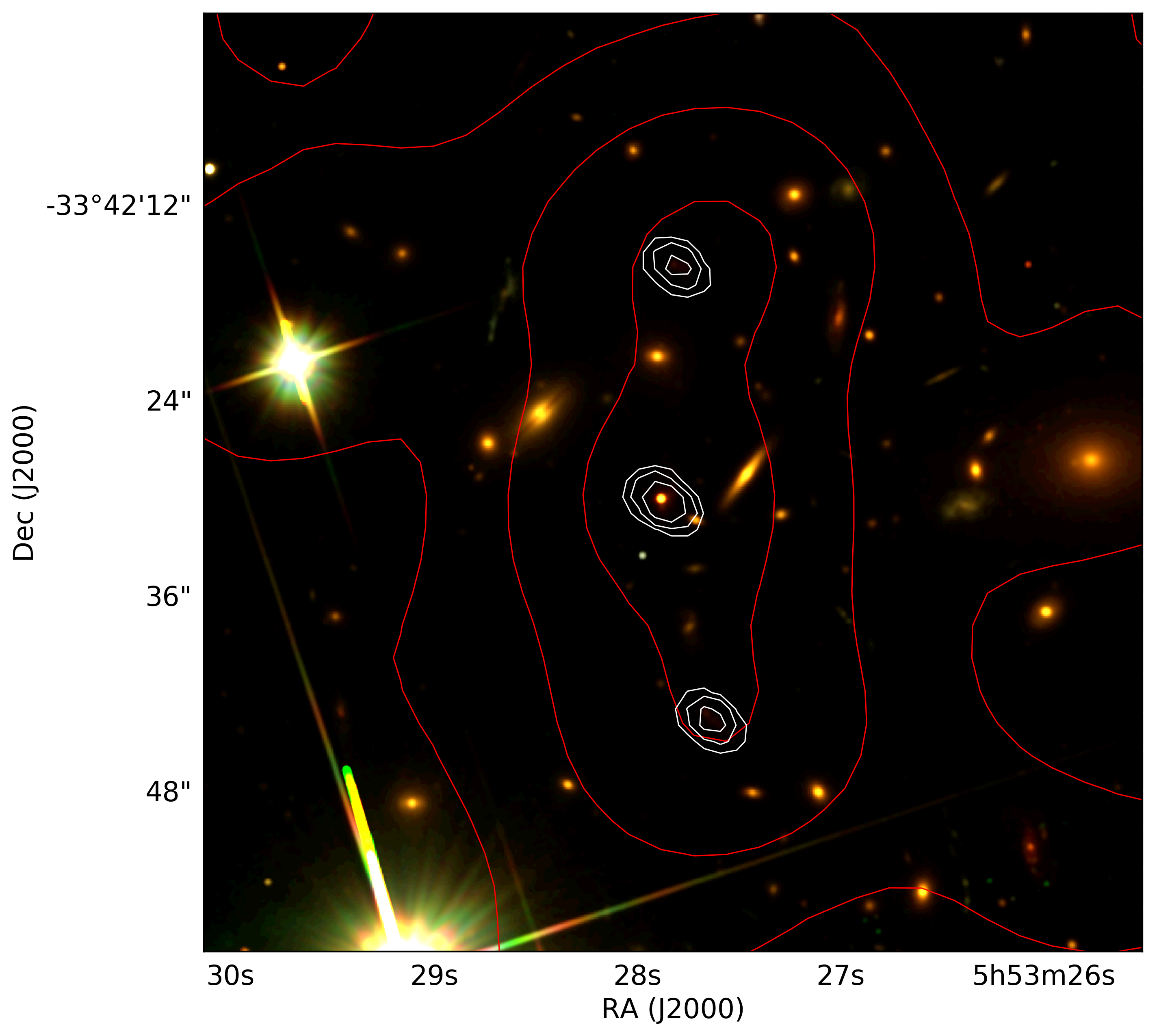}
\caption{Left: Smoothed Chandra X-ray emission of MACSJ0553 with our ASKAP DD image overlaid as contours (levels are same as Fig.~\ref{MACSJ0553opt}: Left). The red region highlights the location of the shock detected in MACSJ0553 and is a reproduction from \citet{2017MNRAS.472.2042P}. Right: HST {\it i,r,g} image of MACSJ0553 with VLA S-band radio emission and ASKAP 943~MHz radio emission overlaid as contours. VLA emission is shown by white contours $[3, 6, 12, 24]~\times~\sigma$ where $\sigma = 10~\mu$Jy beam$^{-1}$. ASKAP emission is shown by red contours $[6, 12, 24]~\times~\sigma$ where $\sigma = 25~\mu$Jy beam$^{-1}$.    \label{MACSJ0553galaxies}}
\end{figure*}

\subsection{MACSJ0553.4-3342}
In Fig.~\ref{MACSJ0553opt}: Left we present our ASKAP image at 943~MHz after DD calibration, which shows an extended radio halo in MACSJ0553. There also appears to be a N-S elongated patch of radio emission to the East of the cluster centre, near the area where \citet{2017MNRAS.472.2042P} report the presence of a shock. Although somewhat elongated, there does appear to be three individual compact components of this structure that lie close together in a N-S orientation, as seen in projection. In our GMRT 323~MHz image (not shown), this region of emission appears much more arc-shaped\footnote{This is due to elongation of the beam since the GMRT is at a Northern latitude and the observation pointing is at a high Southern declination.}, and it is not possible to distinguish the three cores that are partially visible in the ASKAP data. We were therefore unsure whether this emission was associated with one or more radio galaxies or if it was instead generated by a merger shock, which would make it a radio relic candidate. Although this brightened radio region does not coincide with the brightness edge seen in X-ray (see Fig.~\ref{MACSJ0553galaxies}: Left), mock radio relic simulations have shown that gischt-type relics can form within the X-ray boundary of the ICM \citep{2017MNRAS.470..240N} or appear closer to the cluster center depending on the viewing angle \citep{2013ApJ...765...21S}. This has been seen in a few clusters, e.g. Abell 959 \citep{2019MNRAS.487.4775B}, Abell 2255 \citep{2016A&A...593L...7A}, and MACS J0717.5 + 3745 \citep{2009A&A...503..707B}. 

To confirm whether this emission was coming from one or more radio galaxies, we utilized the high-resolution (configuration A) VLA S-band observations of the cluster. When overlaying our VLA image with a resolution of 2.5 arcsec $\times$ 1.2 arcsec on the HST optical image, it is clear that this emission is in fact produced by either three separate compact radio galaxies or by one radio galaxy with two hotspots (see Fig.~\ref{MACSJ0553galaxies}: Right). As an estimate from the colour and size of the optical counterparts, the source in the middle appears to belong to a resident galaxy of the cluster, while the Northern and Southern radio sources appear to be associated with faint background galaxies, however, the N and S sources could possibly be equidistant hotspots from the AGN of the middle, resident galaxy. There are no reported redshifts available for these galaxies. 

To measure the integrated flux density of the radio halo in MACSJ0553, we performed a point source subtraction of the three embedded AGN. Our method for subtraction is described in Sect.~\ref{sub}. In the full bandwidth AKSAP image, it is apparent that there are some additional, very faint point sources around the border of the halo emission. Two of these very faint, compact sources appear to coincide with disk-hosting galaxies, and may come from very small-scale Seyfert jets. One point source to the East is beyond the edge of the HST optical image, so we do not know whether it belongs to a foreground or background galaxy. However, none of these faint point sources appear in any of our six sub-band images, nor in our GMRT or VLA images, and could not be measured with the {\sc Aegean} \citep{2012MNRAS.422.1812H, 2018PASA...35...11H} source detection software, so they were not modeled or subtracted. 

In Fig.~\ref{MACSJ0553opt}: Right we show the radio halo emission imaged after subtracting the three N-S compact AGN and subsequent DD calibration. To capture emission on larger scales, this image was made with a Brigg's robust setting of -0.75 and a restoring beam of 20 arcsec. We measure the integrated flux density of the radio halo within a polygon region marked by the $3\sigma$ contour line\footnote{The polygon is drawn such that the bordering point sources to the East and South are not included.}, where $\sigma = 25~\mu$ Jy per beam. The halo has a flux density of $S_{943}~{\rm MHz} = 12.22 \pm 1.37$~mJy and the largest linear size (LLS) is 0.9 Mpc. The error is calculated from the estimated error on the flux scale ($10\%$) and from the error in measuring the flux density when considering the rms noise ($\pm~0.17$~mJy). The halo traces the X-ray emission quite well, filling the full inner volume of the cluster; however, it is slightly more elongated from East to West and shortened from North to South. 

We also derive the integrated spectral index estimate of the radio halo by comparing the diffuse emission (after point source subtraction) as seen by the GMRT at 323 MHz to the diffuse emission as seen by ASKAP. To do this we made images with uniform weighting and the same minimum $uv$ range of $100~\lambda$, regridded the GMRT image to the ASKAP image, and smoothed both of the images to the same beam size (22 arcsec). From these images we measure the flux density within the same region, tracing the $3\sigma$ contour line of the 943 MHz AKSAP image. We find that the radio halo has a flux density of $7.61 \pm 0.87$~mJy at 943~MHz and $22.02 \pm 0.92$~mJy at 323~MHz in these uniform-weighted images, and therefore we estimate the spectral index of the halo to be $\alpha_{323}^{943} = -0.99 \pm 0.12$. 

\subsection{Abell S0592}

\begin{figure*}
\centering
\includegraphics[width=0.49\textwidth]{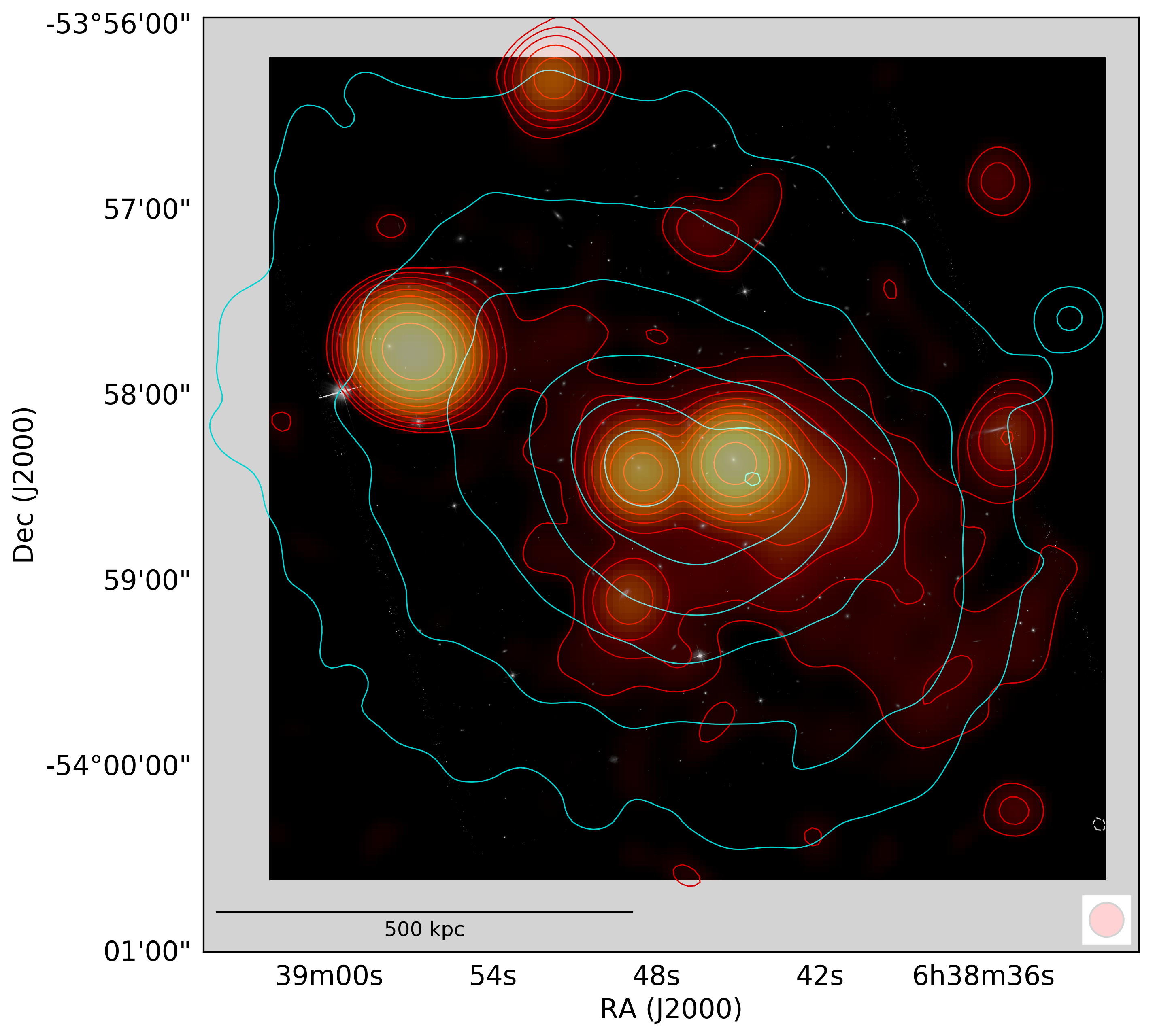}
\includegraphics[width=0.49\textwidth]{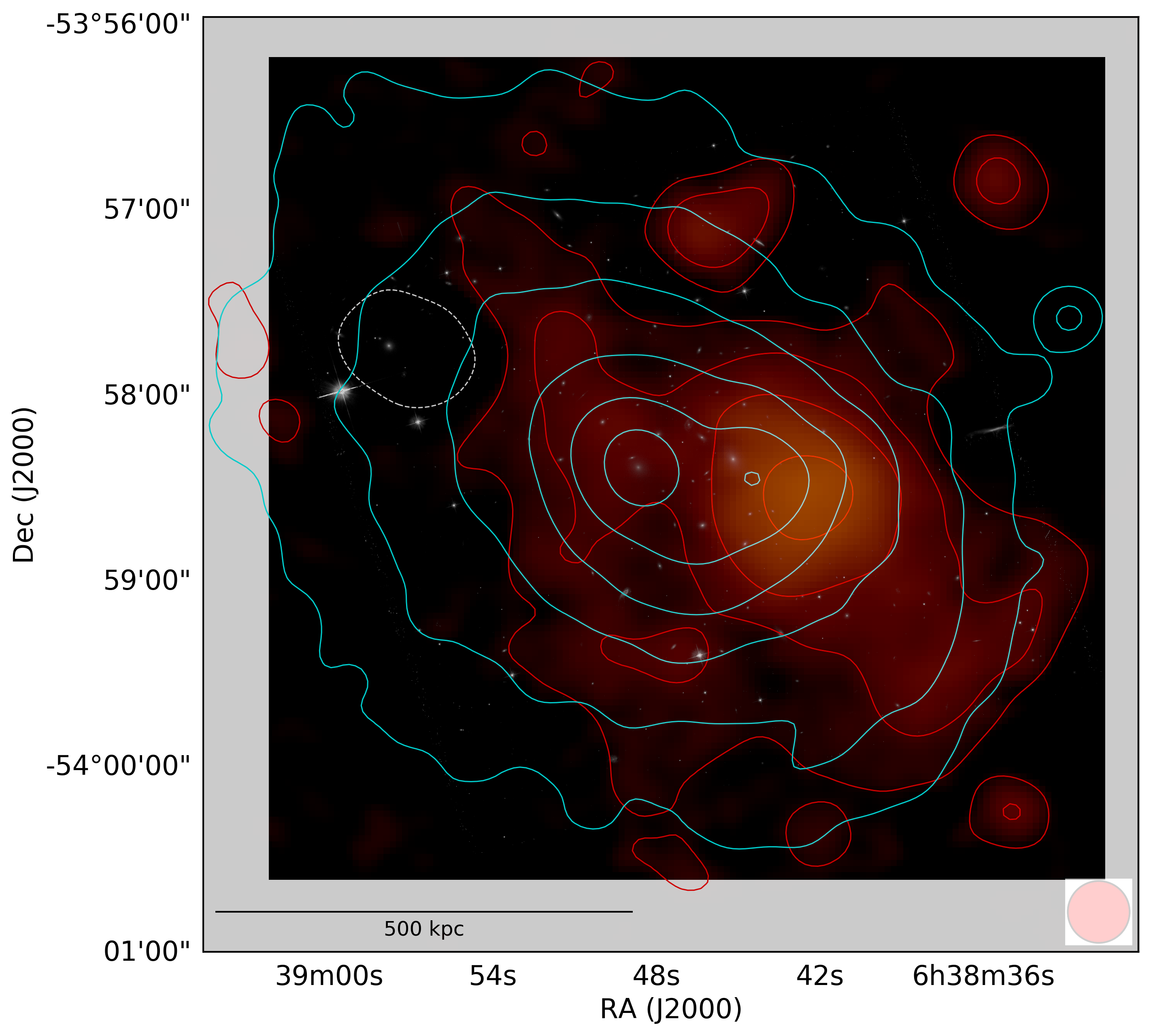}
\caption{HST {\it r} image of AS0592 with ASKAP 1013~MHz radio emission and Chandra X-ray emission overlaid as contours. Left: our final image made with {\sc DDF} ($\sigma = 20~\mu$Jy beam$^{-1}$, restoring beam 11 arcsec $\times$ 11 arcsec). Right: diffuse emission after subtracting compact emission imaged with a $uv$range $>1$km ($\sigma = 25~\mu$Jy beam$^{-1}$, restoring beam 20 arcsec $\times$ 20 arcsec). ASKAP emission in both images is shown by red contours at $[3, 6, 12, 24, 48, 96]~\times~\sigma$ and white dashed contours at $[-2]~\times~\sigma$. Smoothed Chandra X-ray contours as also shown in cyan. The red colour is for visualization only. See text for imaging parameters.   \label{AS0592opt}}
\end{figure*}

Our ASKAP image at 1013~MHz after DD calibration of Abell S0592 (see Fig.~\ref{AS0592opt}: Left) shows diffuse intracluster emission that has not been previously reported in literature. There are four bright, and somewhat extended, radio galaxies embedded in the diffuse emission. One of these radio galaxies, at the cluster center, has a slightly extended lobe that points toward the West. This more extended counterpart of the central radio galaxy appears to bleed into the diffuse emission in the ICM. 

Our image after modeling and subtracting the bright cluster radio galaxies is shown on the right in Fig.~\ref{AS0592opt}. To better capture diffuse emission, this image was made with a Brigg's robust setting of -0.75 and a restoring beam of 20 arcsec. As explained in Sects.~\ref{sub} and ~\ref{subDDF}, we attempted several techniques to properly subtract the radio galaxy emission in this cluster, and found that the best result was to model and subtract the emission using the {\sc DDF} imager after applying DD calibration solutions. The resulting image does leave a hole (negative artefact, marked by the dashed white contour line in Fig.~\ref{AS0592opt}: Right) where the brightest radio galaxy, to the North-East, was subtracted. The diffuse intracluster emission faintly extends within this North-East region of the cluster, so we expect that our flux density measurement in this region will be an underestimation. However, the central radio galaxy with the extended Western lobe appears to leave some residual emission after subtraction, so the flux density measurement in this region will likely be an overestimation. It is difficult to quantify the amount of error that this imperfect subtraction introduces, but we assume a liberal estimate that it is on the order of $\sim~15\%$.  

Due to its LLS of 1.04 Mpc, we classify this diffuse emission as a giant radio halo. From our source-subtracted DD ASKAP image, we measure the flux density of the radio halo within a region marked by the $3\sigma$ contour line where $\sigma = 25~\mu$Jy per beam. We find that the integrated flux density is $S_{1013~{\rm MHz}} = 9.95 \pm 2.16$ mJy. The error is calculated from the estimated error on the flux scale ($5\%$), the error in measuring the flux density when considering the rms noise ($\pm~0.18$~mJy), and an estimate of error due to imperfect source subtraction ($15\%$). 

We were also able to image some diffuse emission in this cluster with ATCA observations. As some of the cluster radio galaxies showed small-scale extended structure in the ATCA data, we imaged the data initially using a balanced Briggs weighting with the robust parameter set to 0.0. No large-scale, diffuse emission was modelled at this stage, and the CLEAN components were subtracted before re-imaging with a natural visibility weighting. A natural weighting was required to maximise the sensitivity to the extended diffuse structure barely significant in the robust 0.0 residuals. The residuals in this naturally weighted image coincide with the radio halo detected in the ASKAP image. The residual emission after source-subtraction is only significant in the full-bandwidth image centered at 2.215~GHz. The flux density of the diffuse emission measured within a region tracing the $2\sigma$ contour line, where $\sigma = 120~\mu$Jy per beam, is $S_{2215}~{\rm MHz} = 3.8 \pm 0.4$~mJy. The error on the flux comes from a $\sim~2\%$ error due to calibration of ATCA data for the 16~cm band and the error from the rms. We are unable to quantify the error introduced from source-subtraction. 

We derive an integrated spectral index estimate of the radio halo in AS0592 by comparing the diffuse emission as seen by ASKAP to the diffuse emission as seen by ATCA. The spectral index estimate was calculated using measurements from our source-subtracted DD-calibrated ASKAP image and our natural-weight ATCA image. Since the spectral index should ideally be calculated from flux density measurements that are taken from images made with uniform weighting, we expect that our value will be biased, but we could not capture any diffuse emission by imaging the ATCA data with a uniform weight. The spectral index of the halo is estimated to be $\alpha^{2215}_{1013} = -1.23 \pm 0.24$. The error on this value comes from the error on the flux density alone, since we are unable to quantify the error due to the different weighting schemes.

\begin{figure}
\centering
\includegraphics[width=0.49\textwidth]{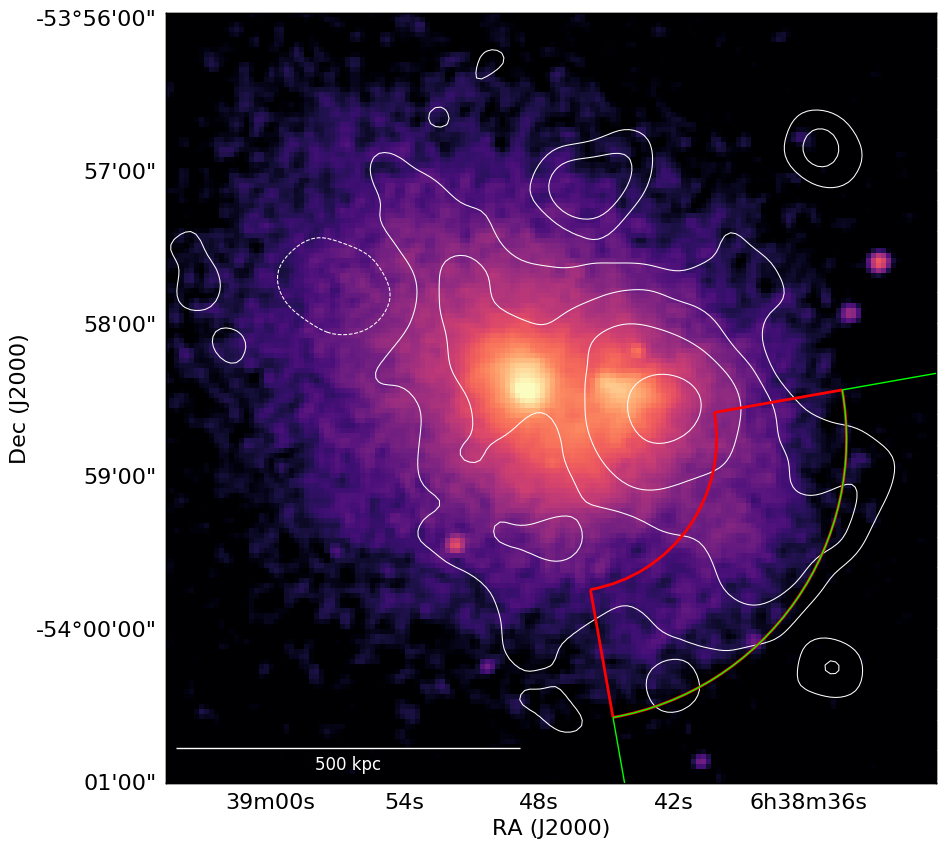}
\caption{Smoothed Chandra X-ray emission of AS0592 with our 1013~MHz ASKAP DD image, after source subtraction, overlaid as contours (levels are same as Fig.~\ref{AS0592opt}: Right.) Panda annulus shows where surface brightness was measured for a radial profile. The yellow curve indicates the SB edge.    \label{AS0592-xray}}
\end{figure}

\begin{figure}
\centering
\includegraphics[width=0.49\textwidth]{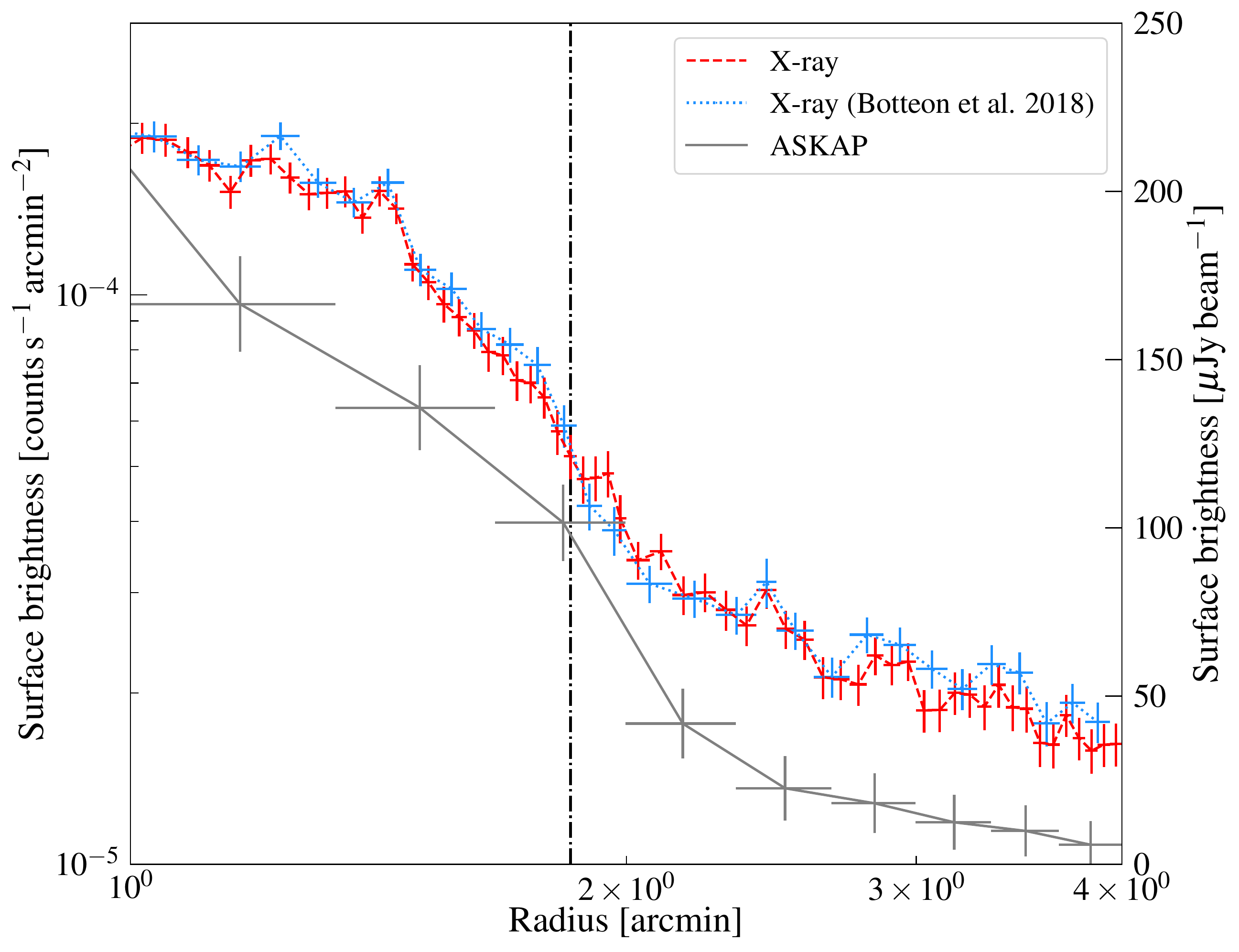}
\caption{The azimuthally averaged SB in our radio AKSAP map is compared to the SB in X-rays. One radio SB measurement is made per beam size (22 arcsec) over 4 arcmin. The dashed vertical line marks the edge corresponding to a jump in SB. Note that radial uncertainties correspond to bin widths. \label{xray-prof}}
\end{figure}

In Fig.~\ref{AS0592-xray} we present our X-ray image of AS0592 from archival Chandra data with our ASKAP DD image (after source subtraction) overlaid as contours. It is apparent that this system has two X-ray peaks with a morphology indicating a Bullet-type merger. The central peak is the brightest, and the second, dimmer peak to the West represents the `bullet.' The radio halo in this cluster is more offset from the X-ray gas, filling the South-Western volume of the ICM, as seen in projection, and it also follows quite closely with the central radio galaxies. The SW border of the radio halo appears to have a more linear edge, roughly coincident with the surface brightness (SB) edge reported by \citet{2018MNRAS.476.5591B}. Because they measured a steep temperature drop across this edge, \citet{2018MNRAS.476.5591B} claim the presence of a shock. We construct a SB radial profile in a wider region\footnote{Region is shown as a panda annulus in Fig.~\ref{AS0592-xray}.} across the SW portion of the cluster and compare it to the profile from \citet{2018MNRAS.476.5591B}, which was measured over a narrower region. In Fig.~\ref{xray-prof} the X-ray SB profiles, constructed using the \texttt{proffit} software \citep{Eckert2011}, are compared to the radio SB profile over the same region. The radio SB profile is computed by azimuthally averaging the SB in radially equal bins equivalent to the size of the synthesized beam (22~arcsec). Uncertainties in the azimuthally averaged surface brightness are estimated via $\langle\sigma_\mathrm{rms}\rangle/\sqrt{N_\mathrm{beam}}$, where $\langle\sigma_\mathrm{rms}\rangle$ is the mean image rms in the bin, and $N_\mathrm{beam}$ is the number of independent beams covering the bin. In Fig.~\ref{xray-prof} it is apparent that the azimuthally averaged SB in radio falls off at a similar rate to the X-ray SB. 

\section{DISCUSSION}\label{sec:discussion}

\subsection{DD calibration for ASKAP}

The results of the DD calibration are noticeable and much more apparent for the brightest sources in the field. In Fig.~\ref{DD} there appears to be some slight ring-type artefacts leftover in each DD image. It is likely that these artefacts are related to imperfect gains calibration in the pre-calibrated measurement sets. The more significant radial stripes and rings, however, have been eliminated through DD calibration. It is worth mentioning how the artefacts in the ASKAPsoft image of IC 5063 are affecting the extended lobes of a nearby radio galaxy, as shown in Fig.~\ref{DDe}. After DD calibration, artefacts are considerably reduced and the diffuse emission of the radio lobes are better recovered, as shown in Fig.~\ref{DDf}. For the three images in Fig.~\ref{DD}, the noise near the bright source is reduced by a factor of two, on average.

To understand why 3GC is effective for ASKAP, we must consider potential sources of directional errors. In general, visibility data from radio interferometers are often affected by baseline-time-frequency direction-dependent effects, which can include (but are not limited to) antenna pointing errors, complex beam shapes, and distortions induced by the ionosphere. Lower frequency radio observations are more severely effected by the highly variable ionosphere (with a plasma frequency emitting around $\sim~10$~MHz). Some progress has been made in modeling the spatial TEC (total electron content) differential for very low frequency observations with LOFAR \citep{2018A&A...615A.179D}, also known as TEC screening. At the higher operating frequencies of ASKAP, the ionosphere is less likely to be the dominating factor in directional errors. 

ASKAP is unlike any other telescope in how its beams are formed. The novel PAFs on ASKAP antennas consist of individual voltage amplifiers that are cross-correlated to form beams digitally. A specific beam ``$j$'' is acquired by applying weights to the elements in the PAF in order to maximise signal-to-noise in that direction. The weights used to form beam $j$ are calculated through observations of the Sun, and can vary from antenna to antenna \citep{2014PASA...31...41H}. The voltages recorded by the elements that form beam $j$ are then cross-correlated between all antennas to produce visibility data in the direction of beam $j$. Beam patterns are revealed in holographic maps (see, for example, Fig. 7 of \citealp{2016PASA...33...42M} and Fig. 10 of \citealp{2019PASA...36....9J}), and appear more eccentric on the edges of the field of view due to coma distortion. However, since the primary beams are modeled as circular Gaussians (with FWHM of $1.09\lambda / D$), primary beam correction does not correct for complex beam shapes. It is likely that imperfections in beamforming and beam modeling are responsible for most of the directional errors present in ASKAP data.

Advancements in simulating and modeling the ionosphere and beam patterns during an observation will allow for much improved directional calibration; however, the faceting technique \citep[e.g.][]{2016ApJS..223....2V, 2018A&A...611A..87T} is currently the best method for tackling direction dependent errors. The \href{https://github.com/mhardcastle/ddf-pipeline}{{\sc DDF} pipeline} \citep{2019A&A...622A...1S} utilises the 3GC software {\sc kMS} and {\sc DDF} and has been constructed to calibrate and image high-band antenna survey observations from LOFAR. Given the results of our tests on ASKAP early science data, we make the argument that our DD ASKAP pipeline can be used to appreciably improve the data products of future survey observations made with ASKAP, albeit more computationally expensive\footnote{The processing of one 22 GB measurement set typically requires 256 GB of RAM and .}. 

\subsection{Subtraction with {\sc DDF}}\label{subDDF}
Subtracting the bright and more extended sources in AS0592 proved to be more of a challenge than subtracting the point sources in MACSJ0553. Most methods of source subtraction involve modeling sources from a source detection software (e.g. PYBDSF \citep{2015ascl.soft02007M} or {\sc Aegean} \citep{2018PASA...35...11H}) or from the CLEAN components of an image. The brighter AGN in AS0592 were causing slight artefacts that appeared as rings in the ASKAPsoft image. Modeling and subtracting these sources prior to DD calibration would leave those ring-type artefacts in the final image, as well as remove potential directional calibrators for this region of the sky. By modifying our DD pipeline and inserting a customised mask, we were able to make an image of the compact AGN emission only and model  and subtract the CLEAN components of that image from the pre-calibrated visibilites. We then continued to run {\sc DDF} once more on the new subtracted data column, while applying the same directional calibration solutions, this time including all baselines above 60 meters and including a customized mask to cover diffuse emission on the scale of the galaxy cluster. The resulting image is the remaining diffuse emission, after modeling and subtracting the compact AGN with directional calibration applied. Unfortunately, we were unable to prevent a negative artefact from occurring where the brightest radio galaxy was subtracted. We attempted to mitigate this effect by raising the CLEAN threshold on the compact image, but found that too much flux was left remaining post-subtraction. 

The central radio galaxy with an extended Western lobe does appear to have emission on the same scales as the radio halo. By decreasing the minimum uv-range in the compact image, more of this extended component could be modeled and subtracted, but it was our opinion that this could possibly remove halo-related emission as well. Therefore, we only modeled and subtracted emission on scales less than 250 kpc. With only the ASKAP and ATCA observations of this cluster, it is not possible to discern how much flux in our subtracted, diffuse emission image is contributed by the central AGN. When modeling the sources as Gaussians, measuring integrated flux densities over sub-band images, and using {\tt subtrmodel} (as was done for MACSJ0553), the remaining image contained more prominent negative artifacts. Modeling the sources through the {\sc DDF} imager's Predict function yielded the best result. These methods and results proved that our DD pipeline could be easily expanded to complete more specific tasks for science-related analysis, specifically in utlising additional options available in the {\sc DDF} package.

\subsection{On the absence of radio relics}

Although MACSJ0553 is in merging state, after core passage, and a shock and cold front have been detected in X-rays, there is no detectable radio shock emission. The absence of a radio counterpart associated with a confirmed shock has been seen in some other clusters, such as in MACS J0744.9+3927 \citep{2018MNRAS.476.3415W}. Following the method for calculating shock acceleration efficiency as presented in \citet{2018MNRAS.476.3415W} we carry out the same calculations here for the shock detected in MACSJ0553 to determine whether this shock would be able to generate a detectable radio relic.

Taking the parameters of the shock wave as measured by \citet{2017MNRAS.472.2042P} -- a Mach number of $\mathcal{M} = 1.33$ and shock velocity\footnote{We determine the shock velocity by calculating the sound speed in an ideal monoatomic gas using the temperature of the shocked region and the Mach number of the shock as reported in \citet{2017MNRAS.472.2042P}.} $V_{\rm sh}=1892$ km s$^{-1}$ -- and the non-detection of a radio relic, we can compute an upper limit on the particle acceleration efficiency. Comparing the dissipated kinetic power at the shock to the total power in the radio emission\footnote{We use the flux density in the pie cut region shown in Fig.~\ref{MACSJ0553galaxies}.}, we can estimate the acceleration efficiency using equation 2 in \citet{2016MNRAS.460L..84B}:

\begin{equation}
\int_{\nu_{0}} L(\nu) d\nu 	\simeq \frac{1}{2} \eta_{\rm e} \Psi \rho_{\rm u} V_{\rm sh}^{3} (1 - C^{-2}) \frac{B^{2}}{B_{\rm cmb}^{2} + B^{2}} S ,
\end{equation}
where $\eta_{\rm e}$ is the acceleration efficiency, $\rho_{\rm u}$ is the upstream density, $V_{\rm sh}$ is the shock velocity, $C$ is the compression factor which is related to the Mach number via $C = 4 \mathcal{M}^{2} / (\mathcal{M}^{2} + 3)$, $B$ is the magnetic field strength and $B_{\rm cmb} = 3.25(1+z)^2$, $S$ is the surface area of the shock\footnote{This area is the largest linear length times the largest linear width of the shocked region, defined by the pie cut in Fig.~\ref{MACSJ0553galaxies} ($621 \times 212$ kpc$^{2}$).}, and $\Psi$ is the ratio of the energy injected in electrons emitting over the full spectrum versus electrons emitting in radio wavelengths, given by

\begin{equation}
\Psi = \frac{\int_{p_{0}} Q(p) E(p) dp}{\int_{p_{\rm min}} Q(p) E(p) dp} ,
\end{equation}
where $Q(p) \propto p^{-\delta_{\rm inj}}$ and $\delta_{\rm inj} =  2(\mathcal{M}^{2} + 1)/(\mathcal{M}^{2} - 1)$ \citep{1987PhR...154....1B}. The momentum, $p_0$, is the momentum associated with electrons that emit the characteristic frequency of the synchrotron emission, $\nu_0=p_0^2 e B /2\pi m_{\rm e}^3c^3$. Here, $m_{\rm e}$ is the electron mass, $e$ its charge, and $c$ the speed of light. We used the value from Table 5 in \citet{2017MNRAS.472.2042P} for the upstream density in the shocked region ($\rho_{\rm u} = 0.3 \times 10^{-4}~$cm$^{-3}$), and since the magnetic field in this cluster is not known we assume a value of $B=1~\mu$G. For the minimum momentum in the denominator, $p_{\rm min}$, we consider two cases: 1) a low value of $p_{\rm min} = 0.1 m_{\rm e}c$ representing electrons accelerated from the thermal pool, and 2) a higher value of $p_{\rm min} = 100 m_{e}c$ representing a population of relativistic seed electrons. However, we find that in both cases the efficiency has to be unrealistically high, $\gg 100\%$, and we cannot infer an upper bound for $\eta_{\rm e}$. This is likely due to the fact that the upstream density in the shocked region is very low. A relic is not observed, and given these acceleration efficiency calculations a relic would not be expected to form.

The clear Bullet-type merging cluster AS0592 also appears to host a shock, as measured by \citet{2018MNRAS.476.5591B} and confirmed in our X-ray SB profile. Of the radio emission in this region, there does not appear to be any substantial brightening or structural morphology that resembles a radio relic. Instead the radio halo exhibits a linear edge roughly coincident with the location of the shock. There have been several cases where borders of radio halos are observed to coincide with, or be bounded by, surface brightness edges detected in the thermal X-ray emission \citep[e.g.][]{2005ApJ...627..733M, 2011MNRAS.412....2B, 2014A&A...561A..52V, 2014MNRAS.440.2901S, 2016ApJ...818..204V, 2018ApJ...856..162W}. Compression from the shock in this region would only be confirmed through polarization measurements and high-resolution spectral maps. As posed by \citet{2019SSRv..215...16V} there could be turbulence after the passage of a shock, such that a previously formed radio relic now appears blended with the halo emission. We cannot compute the acceleration efficiency of the shock because it is not possible to define a region of radio emission that is potentially generated by the shock. This halo-shock connection in AS0592 is very similar to the case in Abell 520, where a Bullet-type shock front is also discovered to be coincident with the SW edge of a halo, although there the radio emission increases more sharply over the shock front from West to East (upstream to downstream) \citep{2019A&A...622A..20H}. 

For both clusters, the shocks detected are relatively weak. It has been proven that stronger shocks ( $\mathcal{M} = 3-4$) are typically necessary to produce observable radio relics \citep{2014ApJ...785..133H}. With the results of this paper we confirm two more cases where merger-induced shocks do not have clear radio counterparts. 

\begin{figure}[h]
\centering
\includegraphics[width=0.49\textwidth, trim=0 0 0 1.5cm clip]{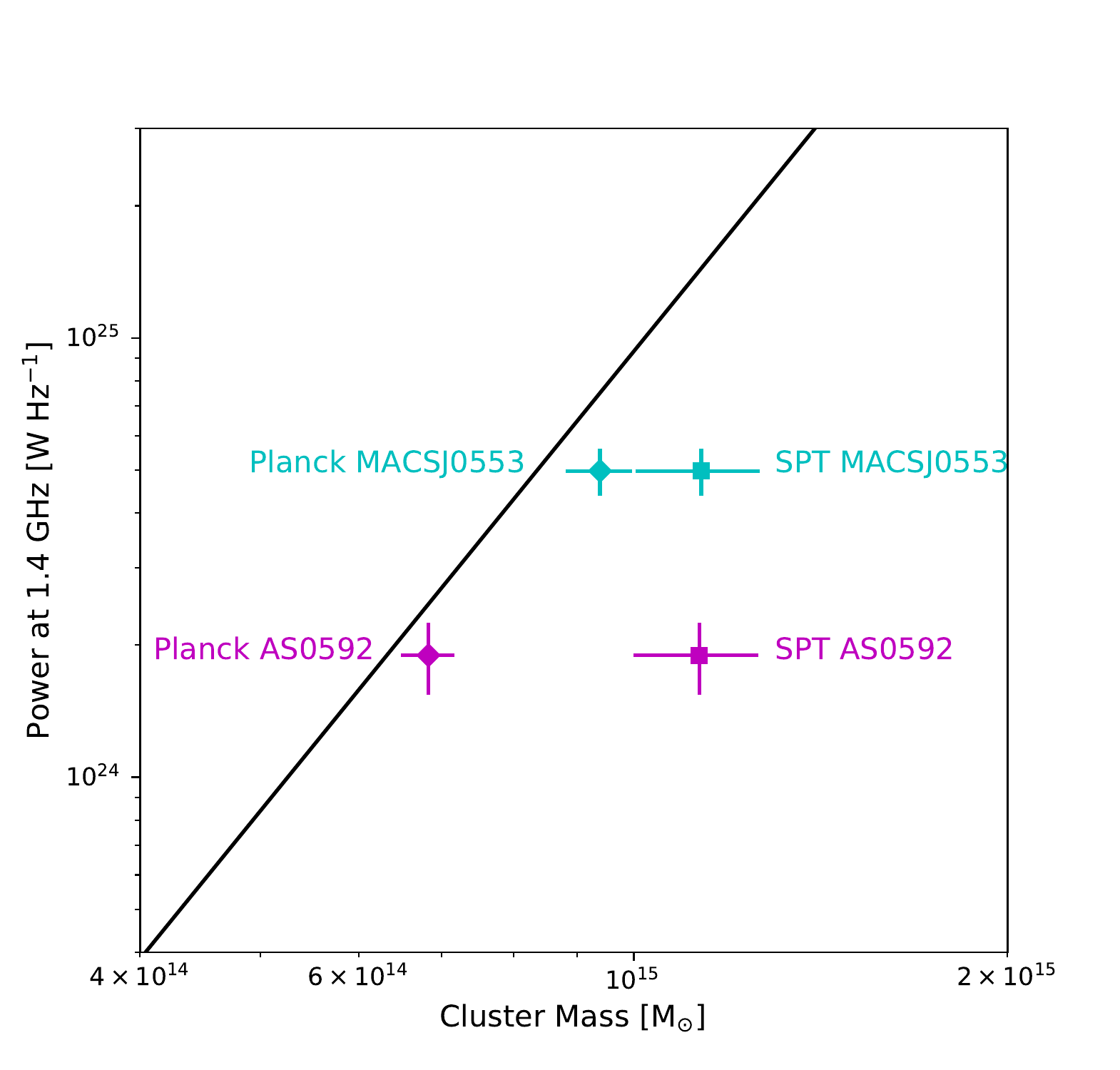}
\caption{The power of the halos in MACSJ053 and AS0592 are extrapolated to 1.4 GHz and plotted against their differing mass estimates from SPT and Planck. The derived fit, or $P-M$ correlation, for a sample of halos with flux measured at 1.4 GHz is shown as a black line, from \citet{2016A&A...595A.116M}.\label{halo-corr}}
\end{figure}

\subsection{On the origins of the radio halos}

Although the masses estimated from SPT observations of the two clusters are relatively similar, the Planck estimated mass of these clusters differ substantially, with a larger discrepancy for AS0592: the mass estimate from SPT observations ($M_{500} = 11.29^{+1.36}_{-1.10} \times 10^{14}$ M$_{\odot}$) is almost twice the mass estimated from Planck observations ($M_{500} = 6.83^{+0.34}_{-0.31} \times 10^{14}$ M$_{\odot}$). In a study of mass calibration for SPT observations, \citet{2015ApJ...799..214B} found that the average cluster masses in a catalogue of 100 SPT clusters were consistently greater (by $\sim~32\%$) than their previous study \citep{2013ApJ...763..127R}, likely due to updated cosmological data. However, there are no published studies explicitly addressing discrepancies between Planck and SPT cluster masses. 

In Fig.~\ref{halo-corr} we plot the halos in MACSJ0553 and AS0592 by their power at 1.4 GHz\footnote{Extrapolated to 1.4 GHz from our measured ASKAP flux densities using our spectral index estimates for each halo, and including a $k$-correction. } versus their cluster mass as listed by both the SPT and Planck catalogues, and compare them to the $P-M$ correlation reproduced from \citet{2016A&A...595A.116M}. In this plot it is easy to see that although the SPT masses of the clusters are very similar, their radio powers are very different, with the halo in MACSJ0553 being much more luminous. Given the SPT mass estimates, both of the halos also lie far outside of the current $P-M$ correlation at 1.4 GHz. If, however, one considers the Planck estimated mass, the halo powers agree more closely with the correlation that suggests that a lower mass cluster will host a lower luminosity radio halo. 

In an evolutionary simulation, \citet{2013MNRAS.429.3564D} found that the X-ray luminosity of a merging cluster will change over the lifetime of the merger, and that the power emitted by a merger-generated radio halo is transient, rising and falling along the $P-L$ correlation. This has two interesting implications: 1) because of its transience, X-ray luminosity is not an entirely reliable property for measuring cluster mass, and 2) if AS0592 and MACS0533 have similar masses, as indicated by the SPT-SZ measurements, the differences in the halo powers may be connected to the evolutionary state of the mergers. 

As seen in our DD AKSAP image, the radio halo in MACSJ0553 fills the ICM and traces the X-ray emission of the cluster very well, indicating that the full volume of the ICM contains ultra-relativistic electrons. Except for the small point-source AGN coinciding with a single cluster galaxy, there does not appear to be any other radio emitting galaxies in the cluster environment. This begs the question of the origin of such a uniform population of CRe in this cluster. A calculation made by \citet{2012MNRAS.426...40B} was that the timescale after core passage would not allow sufficient time for a halo of that size to generate from turbulent re-acceleration alone. These clues lend support to the hadronic model, in which ultra-relativistic electrons can be generated by the collisions of thermal and cosmic-ray protons in the ICM. The luminosity of the halo and the flatter spectral index are also more in line with the hadronic model. It would be interesting to look at sensitive $\gamma$-ray observations of this cluster to determine whether its $\gamma$-luminosity is above the average upper limit of other clusters.

In contrast, AS0592 contains several bright radio galaxies that are injecting relativistic electrons into the surrounding environment. The radio halo in AS0592 tightly hugs the regions where the radio galaxies are present, and the central Western radio galaxy even appears to have lobe emission that bleeds into the emission of the halo. Given its morphology and proximity to the cluster radio galaxies, it is probable that the radio halo was generated from a contributed population of AGN seed, or remnant, electrons, which have then been `boosted' by the turbulence of the cluster merger. We suspect that observations from the upcoming EMU survey will reveal many more examples where diffuse cluster sources are clearly fed by remnant AGN emission.

The turbulent re-acceleration model predicts that radio halos will be generated by a larger population of lower-energy electrons, yielding a steeper spectral index \citep{2008Natur.455..944B}. Our spectral index estimate does not indicate an ultra-steep halo, but we are wary of regarding this value as accurate because it was difficult to recover the diffuse emission in this cluster with higher frequencies. We were also unable to utilise low-frequency MWA observations due to the lower resolution of the interferometer and significant source-blending, as seen in GLEAM images. High-resolution, low-frequency observations would be needed to accurately subtract the compact radio galaxies and measure the flux density of the radio halo in this cluster. Nonetheless, due to its more irregular morphology, lower luminosity, and association with nearby galaxies, we argue that the halo in AS0592 falls more in line with the turbulent re-acceleration model.

\section{CONCLUSIONS}\label{sec:conclusions}

Early science survey observations from ASKAP were made public through CASDA in 2019. Examining the data we found signs of diffuse emission associated with two high-mass, merging clusters: MACSJ0553 and AS0592. We pulled the pre-calibrated data sets covering these clusters from CASDA and performed further processing using the {\sc DDFacet} and {\sc killMS} \citep{2018A&A...611A..87T} software to calibrate against directional errors in the data, greatly improving upon the images available on CASDA -- especially for regions of the sky where bright radio sources cause strong artefacts.

We find that for bright sources causing radial ring-type or stripe-type artefacts, directional calibration reduces the noise by a factor of two within 10 arcmin of the source. Artefacts from bright radio galaxies within and near the clusters were also eliminated through DD calibration. In considering these results, we argue that our DD pipeline can be used to create value-added data products for survey observations with ASKAP.

From our DD ASKAP images we confirm the presence of a giant radio halo in MACSJ0553 (previously detected by \citealp{2012MNRAS.426...40B}) and announce the discovery of a giant radio halo in AS0592. The halos have a similar LLS of $\sim~ 1~$Mpc, but differ in radio luminosity despite the similar masses of the clusters. We find the flux densities of the halos to be $S_{943}~{\rm MHz} = 12.22 \pm 1.37$~mJy and $S_{1013~{\rm MHz}} = 9.95 \pm 2.16$ mJy for MACSJ0553 and AS0592, respectively. In comparing measurements at other frequencies, we estimate the integrated spectral indices to be $\alpha_{323}^{943} = -0.99 \pm 0.12$ for the halo in MACSJ0553 and $\alpha^{2215}_{1013} = -1.23 \pm 0.24$ for the halo in AS0592. 

The halo in MACSJ0553 traces the thermal emission of the cluster extremely well and does not appear to be fed by seed electrons from any cluster AGN. Due to its size, morphology, high luminosity, and flatter spectral index, we argue that this halo falls more in line with the hadronic model. In contrast, the radio halo in AS0592 is more irregular in shape and does not fill the full volume of the cluster, but instead closely traces the location of resident radio galaxies. We argue that these cluster radio galaxies contribute a population of seed electrons that have been re-enegerised by the major merger of the cluster. Due to its morphology, lower radio luminosity, its association with radio galaxies in the cluster, and the fact that the halo is offset from the X-ray gas, we argue that this halo falls more in line with the turbulent re-acceleration model. 

As indicated by their thermal X-ray emission, both clusters have independently undergone major mergers and are in the phase after core passage. Both clusters also show SB edges in their X-ray radial profiles, corresponding to shocks as confirmed by a jump in temperature. However, neither of the clusters host a radio shock. While the merger shocks have relatively low mach numbers, the upstream particle density in MACS0553 is too small to produce a radio shock. In AS0592 the shock boundary coincides roughly with the edge of the radio halo. 

It is compelling to consider the differences in the halos for two clusters that have such similar merger dynamics, mass, and X-ray properties. This reinforces our uncertainty regarding the origin of these sources. High-resolution, low-frequency observations of these clusters will help to constrain the spectral indices of the radio halos, and, in conjunction with more sensitive $\gamma$-ray observations, may provide more clarity on their distinctive origins. 

\begin{acknowledgements}
The Australian SKA Pathfinder is part of the Australia Telescope National Facility which is managed by CSIRO. Operation of ASKAP is funded by the Australian Government with support from the National Collaborative Research Infrastructure Strategy. ASKAP uses the resources of the Pawsey Supercomputing Centre. Establishment of ASKAP, the Murchison Radio-astronomy Observatory and the Pawsey Supercomputing Centre are initiatives of the Australian Government, with support from the Government of Western Australia and the Science and Industry Endowment Fund. We acknowledge the Wajarri Yamatji people as the traditional owners of the Observatory site. The Australia Telescope Compact Array is part of the Australia Telescope National Facility which is funded by the Australian Government for operation as a National Facility managed by CSIRO. This paper includes archived data obtained through the Australia Telescope Online Archive (http://atoa.atnf.csiro.au). This work was supported by resources provided by the Pawsey Supercomputing Centre with funding from the Australian Government and the Government of Western Australia. We acknowledge and thank the builders of ASKAPsoft. AGW acknowledges the Hamburger Sternwarte for its computing resources. AGW  thanks Aidan Hotan, Ron Eckers, Dave McConnell, and Marcus Br\"uggen. SWD acknowledges an Australian Government Research Training Program scholarship administered through Curtin University.

\end{acknowledgements}

\bibliographystyle{pasa-mnras}
\bibliography{askap_halos}

\end{document}